\documentclass[
aps, 
prx, 
10pt,  
amssymb, 
amsmath, 
superscriptaddress, 
tightenlines, 
twocolumn, 
notitlepage, 
] {revtex4-2}
\usepackage{graphicx}
\usepackage{amsmath}
\usepackage{amssymb}
\usepackage{bm}
\usepackage{color}
\usepackage{hyperref}
\usepackage{tabularx}
\usepackage{multirow}
\usepackage{booktabs}
\usepackage{braket}
\usepackage{epstopdf}

\begin{document}

\title{Maximally Localized Wannier Functions, Interaction Models and Fractional Quantum Anomalous Hall Effect in Twisted Bilayer MoTe$_2$}

\author{Cheng Xu}
\affiliation{Department of Physics and Astronomy, University of Tennessee, Knoxville, TN 37996, USA}
\affiliation{Department of Physics, Tsinghua University, Beijing 100084, China}

\author{Jiangxu Li}
\affiliation{Department of Physics and Astronomy, University of Tennessee, Knoxville, TN 37996, USA}

\author{Yong Xu}
\affiliation{Department of Physics, Tsinghua University, Beijing 100084, China}

\author{Zhen Bi}
\affiliation{Department of Physics, The Pennsylvania State University, University Park, Pennsylvania 16802, USA}

\author{Yang Zhang}
\affiliation{Department of Physics and Astronomy, University of Tennessee, Knoxville, TN 37996, USA}
\affiliation{Min H. Kao Department of Electrical Engineering and Computer Science, University of Tennessee, Knoxville, Tennessee 37996, USA}

\begin{abstract}
We investigate the moir\'e band structures and the strong correlation effects in twisted bilayer MoTe$_2$ for a wide range of twist angles, employing a combination of various techniques. Using large-scale first principles calculations, we pinpoint realistic continuum modeling parameters, subsequently deriving the maximally localized Wannier functions for the top three moir\'e bands. Simplifying our model with reasonable assumptions, we obtain a minimal two-band model, encompassing Coulomb repulsion, correlated hopping, and spin exchange. Our minimal interaction models pave the way for further exploration of the rich many-body physics in twisted MoTe$_2$. Furthermore, we explore the phase diagrams of the system through Hartree-Fock approximation and exact diagonalization. Our two-band exact diagonalization analysis underscores significant band-mixing effects in this system, which enlarge the optimal twist angle for fractional quantum anomalous Hall states.
\end{abstract}
                          	
\maketitle

\section{Introduction}

Transition metal dichalcogenide (TMD) based moir\'e materials have proven to be a fertile ground for unveiling new quantum states of matter \cite{kennes2021moire,mak2022semiconductor}. The inherent large effective mass of TMDs ensures that these moir\'e superlattices consistently present narrow moir\'e bands across a range of twist angles, eliminating the need for fine tuning in experiments. These materials exhibit a rich variety of correlated electron states due to pronounced interaction effects, encompassing Mott and charge-transfer insulators \cite{PhysRevLett.121.026402,PhysRevB.102.201115,regan2020mott,tang2019wse2,wang2020correlated,ghiotto2021quantum,li2021continuous,https://doi.org/10.48550/arxiv.2202.02055}, generalized Wigner crystals \cite{regan2020mott,xu2020correlated,zhou2021bilayer,jin2021stripe,li2021imaging,huang2021correlated,padhi2021generalized,matty2022melting}, and quantum anomalous Hall states \cite{li2021quantum}. Fascinatingly, recent experiments on twisted bilayer MoTe$_2$ have detected clear signatures of both integer and fractional quantum anomalous Hall (QAH and FQAH) states\cite{cai2023signatures}. These observations were made within a range of fairly large twist angles, specifically $\theta\sim 3.3^{\circ}-3.7^{\circ}$, and were confirmed through both compressibility \cite{zeng2023thermodynamic} and transport \cite{park2023observation,xu2023observation} measurements. The fractional quantum anomalous Hall insulator has long been a coveted milestone in condensed matter physics. It not only fundamentally broadens the spectrum of topological phases of matter but also holds promising prospects for harnessing the power of anyons in topological quantum computations at zero magnetic field \cite{parameswaran2013fractional,bergholtz2013topological,neupert2015fractional,liu_recent_2023,RevModPhys.80.1083}. 

Understanding the underlying physics of these correlated phases and determining the ideal condition for their manifestation remains a critical theoretical pursuit. The Hubbard model, along with its extensions that account for extended interactions and spin-orbit coupling, offers a valuable framework to decipher these strongly correlated electronic phases. For instance, in TMD heterobilayers such as WSe$_2$/MoSe$_2$ \cite{wang2019optical}, the triangular lattice Hubbard model provides a natural description of the metal-insulator transition at half filling as holes residing on the WSe$_2$ layer experience a triangular lattice potential coming from the moir\'e structure. By incorporating extended interactions, one can elucidate the diverse Wigner crystals seen experimentally at fractional fillings \cite{xu2020correlated}. To facilitit future work on strongly correlated physics in MoTe$_2$ systems, in this work, we construct the maximum localized Wannier functions and minimal Hubbard-like models for the twisted homobilayer MoTe$_2$ system. 

The concept of the fractional quantum anomalous Hall effect in flat band systems has been theoretically proposed for over a decade \cite{PhysRevLett.106.236802,sheng2011fractional,regnault2011fractional,sun2011nearly,sun2011nearly}. In recent times, within the graphene \cite{repellin2020chern,ledwith2020fractional,wilhelm2021interplay,parker2021field,ledwith2022vortexability}  and 
TMD \cite{devakul2021magic,li2021spontaneous,crepel2022anomalous}-based moir\'e system, theoretical predictions have pointed to such an exotic state at partial filling of the topological moir\'e flat band at small twist angle. This state was later experimentally identified, but at a finite magnetic field in magic angle graphene \cite{xie2021fractional}. 
A key analytical foundation of this story is the so-called trace condition for Chern band systems \cite{roy2014band,parameswaran2012fractional,claassen2015position}. This condition is essential for an isolated Chern band to align with a Landau level problem. As we will explore later in our paper, in systems with multiple bands, where the energy separation is either smaller than or on par with the interaction strength—as seen in the band structure of the twisted homobilayer MoTe$_2$ at experimental twist angles—it might be possible that relying solely on the trace condition is not sufficient to accurately predict the FQAH state. While the previous single-band projected exact diagonalization (ED) studies identified the strong FQAH state at flat band regime \cite{li2021spontaneous,crepel2022anomalous,PhysRevB.108.085117,wang2023fractional}, our multi-band projected ED suggests a tendency for FQAH states to appear at higher twist angles. A more careful exploration of multi-band physics, reminiscent of Landau level mixing in quantum Hall problem, is warranted for future studies.

The subsequent sections of this paper are organized as follows. In Sec. \ref{bandandtopology}, we present the moir\'e band structure from the continuum model with parameters fitted from large scale density functional theory calculation. We observe interesting topological transitions of the top three moir\'e bands as function of the twist angle. A crucial point to be noted is that within our calculation the successful experimental observation of FQAH fall out of the optimal condition for the uppermost moir\'e band. For the twist angles that are relevant to current experiments, the top three moir\'e bands together are topological trivial while each individual band exhibits nontrivial Chern number. In Sec. \ref{threebandmodel}, we obtain the maximally localized Wannier functions for the top three moir\'e bands \cite{Koshino2018d,Kang2018,kaushal2023moire}. These Wannier functions show intriguing layer polarization which indicates that their energies are susceptible to gating field. 

While the three-band model offers a comprehensive description for the system, it poses challenges for numerically exact simulations. Given that the strongly correlated physics primarily occur in the uppermost moir\'e band, it is plausible that topological shifts far beneath the Fermi level do not significantly alter the physics in the first band. Bearing this assumption in mind, Sec. \ref{twobandmodel} introduces a disentanglement method to simplify the three-band model into a more concise two-band version. Projecting the Coulomb interaction to the two bands, we obtain a generalized Hubbard model which incorporates correlated hopping and direct spin exchange terms in addition to the ordinary Hubbard interactions. Such direct spin exchange interaction could be responsible to the strong ferromagnetism observed in a wide range of filling factor ($\nu= -0.4\sim -1.2$) in the experiments \cite{cai2023signatures,park2023observation,xu2023observation}. 

Sec. \ref{meanfieldtheory} and \ref{ED} focus on the correlation physics of the system. In Sec. \ref{meanfieldtheory}, we study the phase diagram at filling $\nu=-1$ and $\nu=-2/3$ as function of twist angle and gating field via Hartree-Fock mean-field theory. In Sec. \ref{ED}, we employ exact diagonalization (ED) to study the fractional quantum anomalous Hall effect at filling $\nu=-2/3$. In particular, we project the Coulomb interaction to the top two moir\'e bands and perform momentum space ED. In contrast to the flat band FQAH state from the singe-band approach, the two-band projected ED stabilizes the FQAH phase at large twist angles, with a gap size close to experimental observation. 

\section{moir\'e band structure and topology}
\label{bandandtopology}
\begin{figure}[t]
\includegraphics[width=\columnwidth]{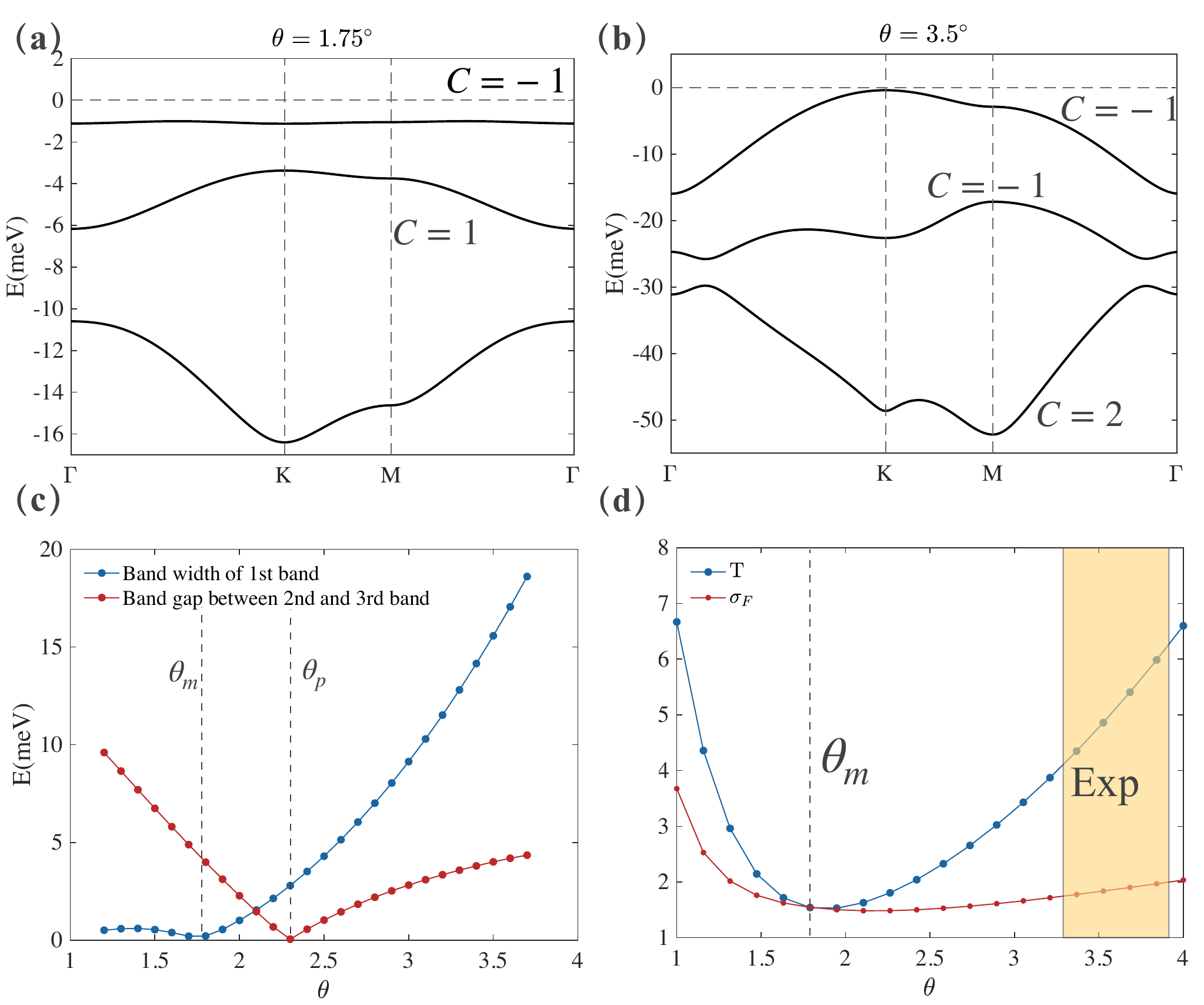}
\caption{(a)/(b) The band structure from continuum model at $\theta=$1.75$^\circ$/3.5$^\circ$.
(c) The red curve is the band gap between the 2$_{nd}$ band and 3$_{rd}$ band as a function of twist angle. The gap closure happens around $\theta_p=2.3^\circ$. The bandwidth of the 1st band is denoted with the blue curve, which attains zero at the magic angle $\theta_m=1.75^\circ$.
(d) $T$ describes the violation of trace condition and $\sigma_F$ reflects the fluctuation of Berry curvature. We show them as a function of twist angles. The optimal condition from these two criteria is around the magic angle $\theta_m$. The orange region denotes the range of angles where FQAH states are observed in experiments.
}\label{fig1}
\end{figure}
The standard treatment for moir\'e band structure is the continuum model \cite{bistritzer2011moire}. The relevant low energy degrees of freedom come from the hole bands from the $K$ and $K'$ valleys of the two layers of MoTe$_2$ \cite{cai2023signatures}. Given that the two valleys are related by time reversal symmetry, it suffices to deduce the band structure from just one valley. For twisted homobilayer TMD systems \cite{wu_topological_2019}, the continuum model for $K$ valley is constrained by rotational and layer-exchange symmetry, and is expressed as:
\begin{equation}
\hat{H}=	\begin{bmatrix}
		-\frac{\boldsymbol{(\hat k-K_{t})^2}}{2m^*}+\Delta_{t}(\boldsymbol{r})& \Delta_{T}(\boldsymbol{r})\\
		\Delta^{\dagger}_{T}(\boldsymbol{r}) & -\frac{\boldsymbol{(\hat k-K_{b})^2}}{2m^*}+\Delta_{b}(\boldsymbol{r})
	\end{bmatrix}	
\end{equation}
with:
\begin{equation}
	\begin{aligned}
		\Delta_{t/b}(\boldsymbol{r})&=2V\sum_{i=1,3,5}\cos(\boldsymbol{g_i\cdot r\pm \phi})\\
		\Delta_{T}(\boldsymbol{r})&=w(1+e^{-i\boldsymbol{g_2\cdot r}}+e^{-i\boldsymbol{g_3\cdot r}})
	\end{aligned}
\end{equation}
where $\boldsymbol{\hat{k}}$ is the momentum measured from the $\Gamma$ point of single layer MoTe$_2$, $\boldsymbol{K_{t}}(\boldsymbol{K_{b}})$ is high symmetry momentum $K$ of the top(bottom) layer, $\Delta_{t}(\boldsymbol{r})(\Delta_{b}(\boldsymbol{r}))$ is the layer dependent moir\'e potential, $\Delta_{T}(\boldsymbol{r})$ is the interlayer tunneling, $\boldsymbol{g_i}$'s are moir\'e reciprocal vectors (geometry of the moir\'e Brillouin zone can be found in Appendix.\ref{fig:Sm_mBZ}).

Here $\Delta_{t/b}(\boldsymbol{r})$ and $\Delta_{T}(\boldsymbol{r})$ are expanded to the lowest order harmonics, which is a standard approximation to make. To obtain accurate parameters in the continuum model, large scale density functional calculations with van der Waals corrections is needed. We have performed a first principle calculation for twisted MoTe$_2$ at twist angle of $\theta=3.89^{\circ}$ (moir\'e wavelength $a_m=5.2$ nm and 1302 atoms) with full lattice relaxation, then fit the moir\'e band structure to obtain the following continuum parameter: $m^*=0.62m_e,V=9.2$ meV, $w=-11.2$ meV, $\phi=-99^\circ$. Using these parameters, we can proceed to solve the moir\'e band structures at arbitrary small angle. 



We can solve the moir\'e band structure with plane-wave basis. For simplicity, we do a standard shift of the momentum origin: $\boldsymbol{k_{t/b}}=\boldsymbol{K_{t/b}-Q_{\pm}+k}$, where $\boldsymbol{k_{t/b}}$ are measured from the $\Gamma$ point of the original Brillouin zone and $\boldsymbol{k}$ is measured from the $\gamma$ point of the moir\'e Brillouin zone. Here  $\boldsymbol{Q_{\pm}=\pm q_1+G_m}$ with $\boldsymbol{q_1}$ being the momentum difference between the $\boldsymbol{K_{t}}$ and $\boldsymbol{K_{b}}$ (and we define $\boldsymbol{q_2}(\bm{q_3})$ as the vector obtained by $C_3(C_3^{-1})$ rotation on $\bm{q_1}$), and $\boldsymbol{G_m}=m\boldsymbol{g_1}+n\boldsymbol{g_3}$ is the moir\'e reciprocal lattice vectors. We also define $\boldsymbol{Q_0=Q_{+}\oplus Q_{-}}$. (More details on the basis can be found in App. VII C 1 
In these plane wave basis, the Hamiltonian reads:
\begin{equation}
	\begin{aligned}
		H&=-\sum_{\boldsymbol{Q\in Q_0}}\frac{(\boldsymbol{k-Q})^2}{2m^*}+\sum_{\boldsymbol{Q,Q^\prime}\in \boldsymbol{Q_0},j=1}^{3}w\delta_{\boldsymbol{Q,Q^\prime\pm q_j}}\\
  &+\sum_{l=\pm}\sum_{\boldsymbol{Q,Q^\prime\in Q_l},i=1,3,5}V(e^{\mathrm{i}l\phi}\delta_{\boldsymbol{Q,Q^\prime+g_i}}+e^{-\mathrm{i}l\phi}\delta_{\boldsymbol{Q,Q^\prime-g_i}})
	\end{aligned}
\end{equation}
where $l=\pm$ is the layer index. Diagonalizing the Hamiltonian, we can obatin the moir\'e band structure. Examples of band structures for $\theta=1.75^\circ$ and $\theta=3.5^\circ$ are shown in Fig. \ref{fig1} (a) and (b). The band width of the uppermost moir\'e band shows a minimal around $\theta\cong 1.75^\circ$ as shown in Fig. \ref{fig1} (c). 

Next we will inspect the topology of these moir\'e bands. For twist angles less than $2.3^\circ$, the Chern numbers, derived from the continnum model for the top three bands, are $-1, 1, 0$. However, for larger twist angles, these Chern numbers transition to $-1, -1, +2$. This observation is further validated using the Wilson loop method (see Appendix\ref{wilsonline} 
for more details). We emphasize that the arrangement of Chern numbers for $\theta>2.3^\circ$ is in agreement with experimental data. So far in all experiments where twist angle $\theta$ ranges between $3.5^\circ-3.9^\circ$, both the Hall conductance and the reflective magnetic circular dichroism increase once the doping exceeds $\nu=-1$. And a double quantum spin Hall effect has been observed at $\nu=-4$. 
These suggest that the second band likely shares the same Chern number as the first band.

Another way to check the Chern number shift is through the $C_3$ eigenvalues of the Bloch wavefunction at high symmetry points in the moir\'e Brillouin zone \cite{zhang2021spin,paul2023giant}. The Bloch wavefunctions at $\kappa, \kappa',\gamma$ are eigenstate of the $\boldsymbol{C_3}$ rotations, namely $\boldsymbol{C_3}|\psi_{n,\boldsymbol{k}}\rangle=\alpha_{n,\boldsymbol{k}}|\psi_{n,\boldsymbol{k}}\rangle$ for $\boldsymbol{k}=\kappa,\kappa',\gamma$. It is known that the Chern number of a given band is related to these three symmetry eigenvalues in the following way \cite{fang2012bulk},
$e^{i 2 \pi C / 3}=\alpha_\gamma \alpha_\kappa \alpha_{\kappa'}$. We have determined the symmetry eigenvalues for the top three moir\'e bands at the high symmetry points (for detailed information, refer to Appendix\ref{dmn}).

When $\theta<2.3^\circ$, the first band $\boldsymbol{C_3}$ symmetry eigenvalues are presented as (1,$e^{-i\frac{2\pi}{3}}$,1) at the points $(\kappa,\gamma,\kappa')$. The second band exhibits symmetry eigenvalues as ($e^{i\frac{2\pi}{3}}$,$e^{-i\frac{2\pi}{3}}$,$e^{i\frac{2\pi}{3}}$), while the third band is represented by (1,1,1). The symmetry eigenvalues align with the Chern numbers of the top three bands, which are $-1,+1,0$. Interestingly, as the twist angle exceeds $2.3^\circ$, we observe a shift in the $C_3$ eigenvalues between the second and third bands at the $\gamma$ point. This suggests a topological transition occurring between these bands. Subsequently, the symmetry eigenvalues correspond with Chern numbers $-1,-1,2$.

We also check the so-called trace condition for the upper most moir\'e band. The band geometry is encoded in the so-called quantum geometry tensor: 	
\begin{equation}
		\eta^{uv}:= A_{BZ}\langle \partial^uu_{\boldsymbol{k}}\mid (1-|u_{\boldsymbol{k}}\rangle \langle u_{\boldsymbol{k}} |)\mid \partial^vu_{\boldsymbol{k}}\rangle
\end{equation}
where $A_{BZ}$ is the area of the Brillouin zone. The symmetric and antisymmetric part of the quantum geometry tensor give the Berry curvature ($\Omega(\boldsymbol{k})=-2\mathrm{Im}(\eta^{xy})$) and quantum metric ($g^{uv}(\boldsymbol{k})= \mathrm{Re}(\eta^{uv})$). To quantify the geometric properties, one can introduce two parameters:
\begin{equation}
	\begin{aligned}
		\sigma_F :& =\bigg[\frac{1}{A_{BZ}} \int d^2{\boldsymbol{k}}(\frac{\Omega(\boldsymbol{k})}{2\pi}-1)^2\bigg]^{\frac{1}{2}}\\
		T:&= \frac{1}{A_{BZ}} \int d^2{\boldsymbol{k}}\bigg[tr(g(\boldsymbol{k}))-\Omega(\boldsymbol{k})|\bigg],
	\end{aligned}
\end{equation}
where $\sigma_F $ describes the fluctuations of Berry curvature and the $T$ quantifies the violation of the trace condition. When both $\sigma_F$ and $T$ tend towards 0, it becomes possible to exactly map the Chern band to a Landau level problem \cite{roy2014band,parameswaran2012fractional,claassen2015position}, allowing for intuitive understanding of the fractional state. We calculate the values of these parameters in relation to the twist angle, as depicted in Fig. \ref{fig1} (d). It's noteworthy that the values of $\sigma_F$ and $T$ are comparable with those identified in magic angle twisted bilayer graphene \cite{ledwith2020fractional,wilhelm2021interplay,parker2021field,ledwith2022vortexability}.

Based on the observations of band topology, we conclude that, for experimentally relevant twist angles, in order to have a localizable Wannier function description, a three-band model is needed. In the following section, we will determine the maximally localized Wannier functions for the top three moir\'e bands. 


\begin{figure*}
\includegraphics[width=2\columnwidth]{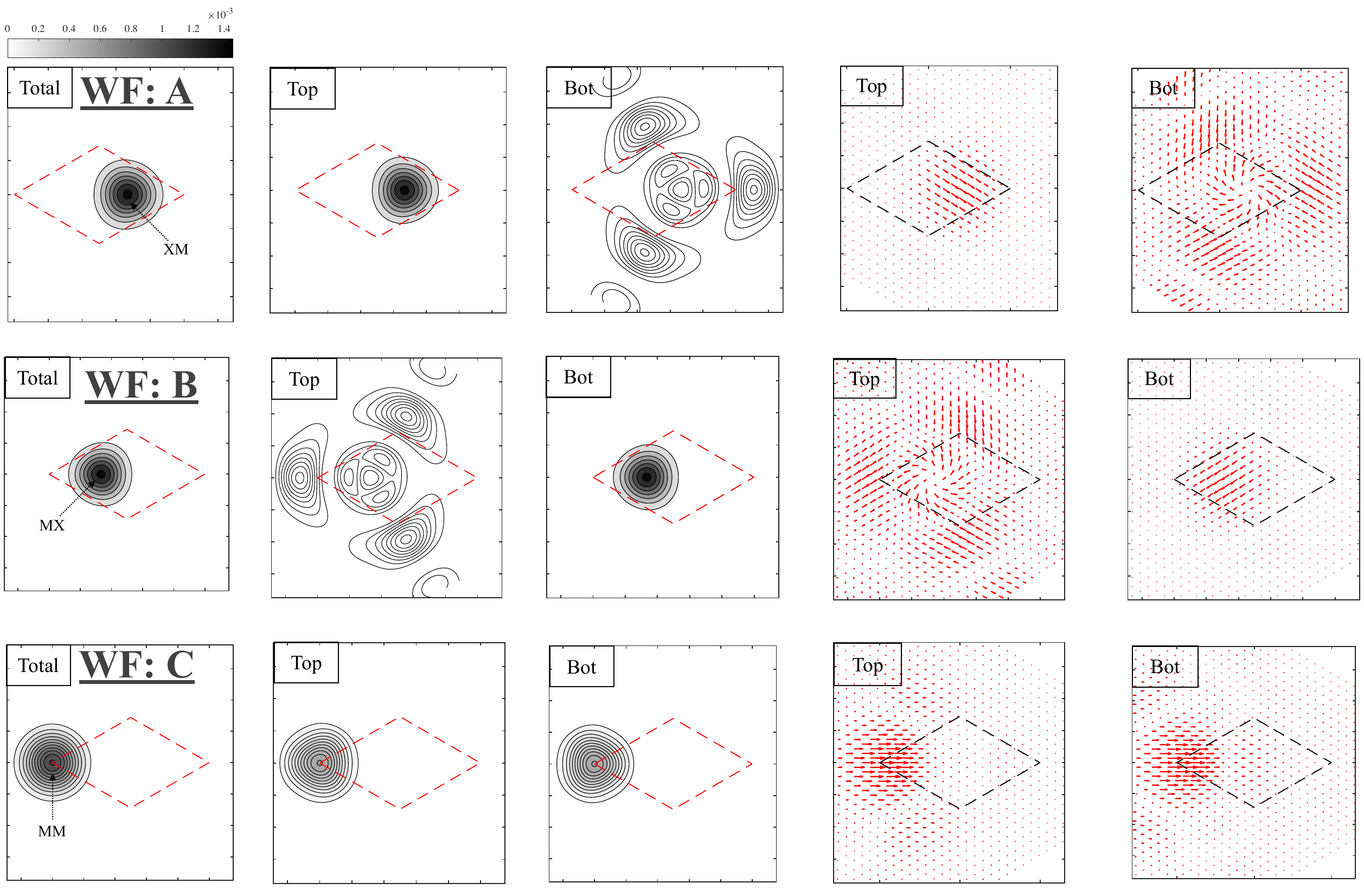}
\caption{The maximally localized Wannier function for the three-band model of twisted bilayer MoTe$_2$ at $\theta=3.5^\circ$. For the three Wannier functions A, B, and C, we demonstrate the total charge distribution (1st column), the charge distributions on each layer (2nd and 3rd columns), and the phase of the wavefunction on each layer (4th and 5th columns). The dashed diamonds are the unit cell of moir\'e superlattice.
}\label{fig2}
\end{figure*}

\section{A three-band model with maximally localized wannier functions}
\label{threebandmodel}
To formulate the maximally localized Wannier functions (MLWF), our initial step is to set the Wannier centers based on the symmetry eigenvalues of the Bloch wavefunctions. There are three Wyckoff positions of the twisted homobilayer TMD moir\'e pattern, namely the MM, MX, and XM stacking regions. The symmetry of the Bloch states at high symmetry momentum is determined by the positions of the Wannier centers. For instance, positioning the Wannier centers at the origin of each unit cell, namely the MM regions of the twisted TMD moir\'e pattern, ensures that the symmetry eigenvalues for the Bloch states will be identity at $\kappa$, $\gamma$  and $\kappa'$. For twist angle smaller than $2.3^\circ$, given the $C_3$ eigenvalues of the third moir\'e band is (1,1,1) at momentum ($\kappa$, $\gamma$, $\kappa'$), its Wannier center should be located at the MM region. For the top two moir\'e bands, to reproduce the correct symmetry eigenvalues, the Wannier centers should be positioned at the MX and XM regions. For a twist angle exceeding $2.3^\circ$, the same set of Wannier functions should remain adequate to characterize the three bands \cite{yu2020giant}. This is because, the topological transition occurs only within this subspace, while the overall Hilbert space for the three bands stays unchanged.

Now we describe the procedure to find the maximally localized Wannier functions. Recall that the Wannier functions are related to the Bloch wavefunction by unitary and Fourier transformations:
\begin{equation}
w_{n\boldsymbol{R}}(\boldsymbol{r})=\frac{1}{\sqrt{N}}\sum_{\boldsymbol{k}}
e^{-i\boldsymbol{k\cdot R}}\sum_{m=1}^3 U_{mn}(\boldsymbol{k})\psi_{m\boldsymbol{k}}(\boldsymbol{r})
\end{equation}
where $\psi_{m\boldsymbol{k}}$ is the Bloch states from the continuum model. $U_{mn}(\boldsymbol{k})$ is a momentum-dependent unitary transformation. $\boldsymbol{R}=m\boldsymbol{a_{m1}}+n\boldsymbol{a_{m2}}$ and $\boldsymbol{a_{m1}},\boldsymbol{a_{m2}}$ are the moir\'e lattice vectors. The approach involves initially proposing a set of trial Wannier functions. Subsequently, by adjusting the parameters in the unitary transformations, we aim to minimize their spread.

We will construct an initial trial function starting from the small twist angle limit. In this limit, we know that the third band is topologically trivial and separated from the first two bands in energy. Therefore, for the third bands, we opt for a gauge where the Bloch states are real and positive at MM regions, ensuring the Wannier functions are anchored at these MM regions. However, for the top two bands, merely selecting a gauge where the Bloch states are real and positive at the Wannier center turns out to not yield an optimal trial Wannier function. 
In addition to the gauge fixing condition mentioned above, we also choose proper $U_{mn}(\boldsymbol{k})$ to maximize the expectation of layer polarize operator\cite{Devakul2021}:
\begin{equation}
	P_{\boldsymbol{k}}=\langle \tilde{\psi}_{n\boldsymbol{k}}|\sigma_z|\tilde{\psi}_{n\boldsymbol{k}}\rangle
\end{equation} 
where $|\tilde{\psi}_{n\boldsymbol{k}}\rangle=\sum_m U_{mn}(\boldsymbol{k})\psi_{m\boldsymbol{k}}$, and the pauli matrix $\sigma_z$ acts in the layer space. These trial Wannier function are well localized and layer polarized and serve as a starting point for searching maximally localzied Wannier functions. 

We search for the unitary transformation that minimizes the spread of Wannier function \cite{Marzari1997b} with the initial $U_{mn}$ constructed above. In this process, we should be careful about the symmetry properties of the Wannier functions \cite{Sakuma2013a}. The symmetry group of the one-valley model is generated by: $C_{2y}\mathcal{T}$ and $C_3$. The WANNIER90 program allows us to minimize the spread of Wannier functions while keeping the $C_3$ symmetry constrain on $U_{mn}(\boldsymbol{k})$ at each step.

The resulted Wannier functions (labeled by A, B, and C) are symmetric under the $C_3$ symmetry. However, the $C_{2y}\mathcal{T}$ symmetry cannot be implemented automatically within WANNIER90. Therefore, at the last few steps, we need to manually restore the $C_{2y}\mathcal{T}$ symmetry. 

Under the $C_{2y}\mathcal{T}$ operation, A and B should be exchanged and C should remain invariant. Empirically, we note that the Wannier functions obtained without accounting for the $C_{2y}\mathcal{T}$ constraint are relatively close to the symmetric limit. To further make the symmetry more valid, we do the following procedure. First, we replace the Wannier function A with the Wannier function obtained by $C_{2y}\mathcal{T}$ operation on the Wannier function B, namely
 \begin{equation}
 		|w_{A}\rangle =C_{2y}\mathcal{T}|w_{B}\rangle
 \end{equation}
For Wannier function C, the $C_{2y}\mathcal{T}$ symmetry generates a constrain on $U_{mn}$:
 \begin{equation}
     U_{mn}(\boldsymbol{k_x,-k_y})=U^{*}_{mn}(\boldsymbol{k_x,k_y})e^{i\alpha_{m}(\boldsymbol{k})},
     \label{c2T}
 \end{equation}
where $\alpha_m$ is defined by $C_{2y}\mathcal{T}| \psi_{m\boldsymbol{k}}\rangle=e^{i\alpha_{m}(\boldsymbol{k})}| \psi_{mC_{2y}\mathcal{T}\boldsymbol{k}}\rangle$. To maintain the $C_{2y}\mathcal{T}$ symmetry, we replace $U_{mn}(\boldsymbol{k_x},\boldsymbol{k_y})$ for $\boldsymbol{k_y}<0$ by the right hand side of Eq. \ref{c2T}. Following these replacements, we re-execute the minimization procedure for the Wannier functions, iterating this method until the symmetry-breaking error is sufficiently reduced. Our computational findings suggest that these substitutions are effective: the band fitting is excellent, and the spread of the Wannier function remains largely unchanged. The profile of the resulting Wannier functions are shown in Fig. \ref{fig2}. 

It is expected that the $\boldsymbol{C_3}$ symmetry eigenvalues of our Wannier function coincide with the $\boldsymbol{C_3}$ symmetry eigenvalues of Bloch states at the $\gamma$ point. This means that the Wannier function A, B, and C should have $\boldsymbol{C_3}$ eigenvalue: $e^\frac{-i2\pi}{3}$, $e^\frac{-i2\pi}{3}$, and $1$. One may be very puzzled about the phase of the Wannier function A and B shown in Fig. \ref{fig2}, where the top and bottom layers seem to transform differently under the $C_3$ symmetry. This is due to the fact that these Wannier functions are plotted with the overall Bloch factor $e^{i\boldsymbol{K_l\cdot r}}$ of the plane wave omitted. Once we carefully account for this factor, the full Wannier functions have the correct symmetry eigenvalues.

The charge distribution of the Wannier function is worth to emphasis. In the whole range of the twist angles from $1^\circ$ to $4^\circ$, the Wannier function A and B is mostly layer polarized. For instance, at $1.2^\circ$, the charge polarization($\int \bigg[ |w^{top}_A(\boldsymbol{r})|^2-|w^{bot}_A(\boldsymbol{r}))|^2\bigg]d\boldsymbol{r}/\int \bigg[ |w^{top}_A(\boldsymbol{r})|^2+|w^{bot}_A(\boldsymbol{r}))|^2\bigg]d\boldsymbol{r}$) is 95.06\%.
The layer polarization will slightly decrease as we increase the twisted angle. However, the charge polarization still remain 88.56\% at 3.7$^\circ$. Due to this property, the displacement field can effectively be modeled as the sublattice potential in the effective tight binding model, which offers a sensitive knob to tune the band structure of the system. On the contrary, the Wannier function C has equal charges on both layers at all angles. It is interesting to note that the charge distribution of top layer component of the Wannier function A is mostly localized in MX region while the bottom layer component has a three-peak structure in the MM regions which resembles the Wannier functions found in twisted bilayer graphene (TBG)\cite{Koshino2018d}\cite{Kang2018}. 

\begin{figure}[t]
\includegraphics[width=\columnwidth]{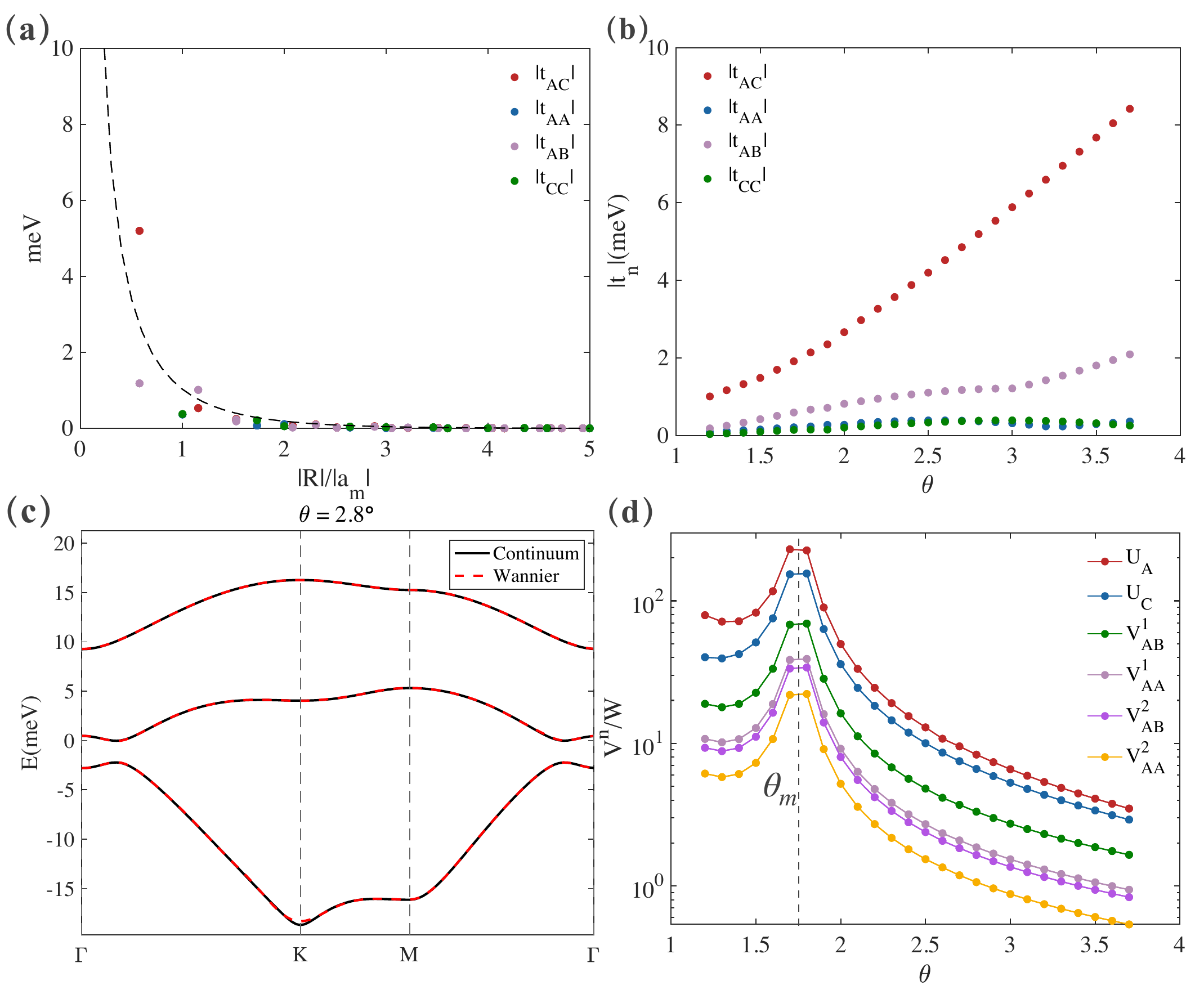}
\caption{(a) The tight-binding parameter as a function of hopping distance at $\theta=2.8^\circ$, and $a_M$ is the lattice constant of moir\'e unit cell. The hopping amplitude decays to zero quickly around $|r|\approx 2|a_{m}|$. The dashed line is a guiding line for view.
(b) The nearest neighbor hopping parameters as a function of twist angle.
(c) The red line is the band structure calculated with our Wannier function , which aligns perfectly with bands from the continuum model(black line). And all of hopping is kept here.($d_{cutoff} \rightarrow \infty$)
(d) The relation between the density-density interaction parameter and the twist angle. $U_A$  is the onsite Hubbard terms, and $V^n_{AB}$ is the $n$-th nearest neighbor term between A and B sites, and so forth. $W$ is the band width of the topmost band. Here we use $\epsilon = 10$ for the dielectric constant and unscreened Coulomb interaction.}
\label{fig3}
\end{figure}

\begin{figure*}[t]
\includegraphics[width=2\columnwidth]{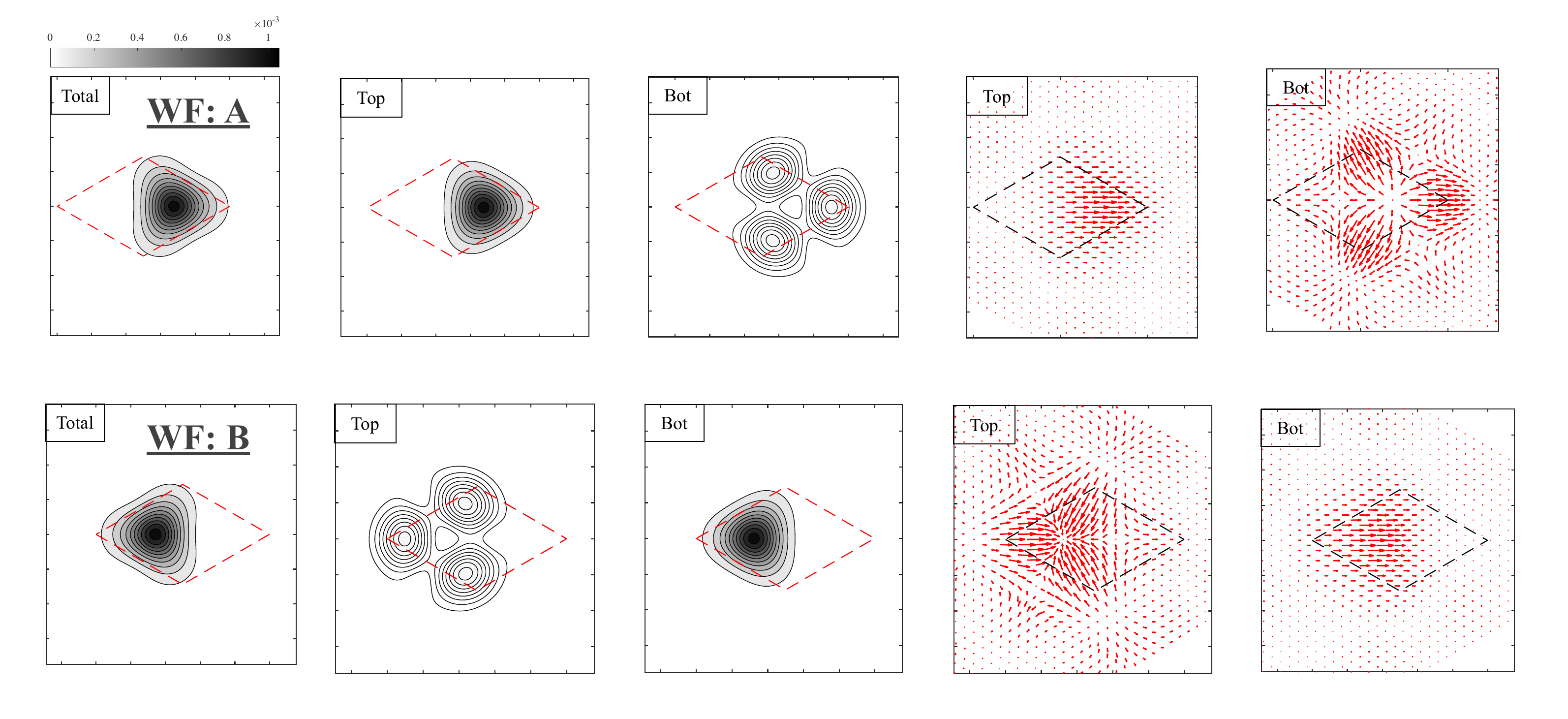}
\caption{The maximally localized Wannier functions for the minimal two-band model of twisted bilayer MoTe$_2$ at $\theta=3.5^\circ$ obtained through the disentanglement procedure. We demonstrate the total charge distribution (1st column), the charge distributions on each layer (2nd and 3rd columns), and the phase of the wavefunction on each layer (4th and 5th columns). The dashed diamonds are the unit cell of moir\'e superlattice.
}\label{fig4}
\end{figure*}

\begin{figure}[t]
\includegraphics[width=\columnwidth]{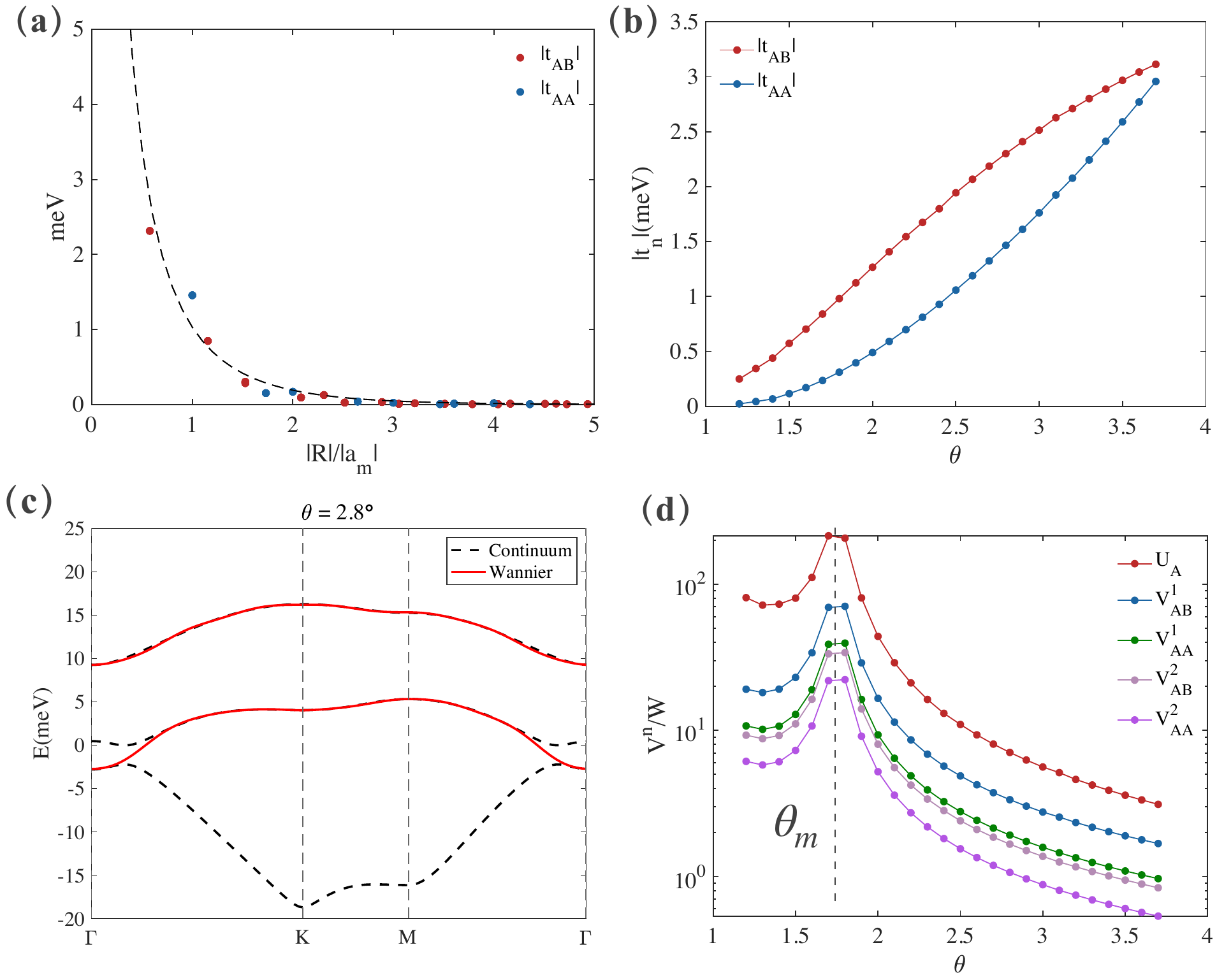}
\caption{(a) The tight-binding parameters for the minimal two-band model as a function of hopping distance at $\theta=2.8^\circ$, and $a_M$ is the lattice constant of moir\'e unit cell. 
(b) The nearest neighbor hopping parameters as a function of twist angle. 
(c) The red line is the band structure calculated with our Wannier functions. The agreement for the top most band with the continuum model calculation (black line) is excellent. And all of hopping is kept here.($d_{cutoff} \rightarrow \infty$)
(d) The relation between the density-density interaction parameter and the twist angle. $U_A$, $U_B$ are the onsite Hubbard terms, and $V^n_{AB}$ is the $n$-th nearest neighbor term between A and B sites, and so forth. $W$ is the band width of the topmost band. Here we use $\epsilon = 10$ for the dielectric constant and unscreened Coulomb interaction.
}\label{fig5}
\end{figure}

Using these Wannier functions, we derive the hopping parameters, which are illustrated in Fig. \ref{fig3} (a). Additionally, the relationship between these hopping parameters and the twist angle is depicted in Fig. \ref{fig3} (b). The hopping terms decrease quickly to zero at about $|r|\approx 2|a_{m}|$, which indicates the Wannier functions are well localized. An exemplar band structure at $2.8^\circ$ utilizing the tight binding model is presented in Fig. \ref{fig3} (c), and it aligns remarkably well with the band structure from the continuum model. The parameters for density-density interaction, in relation to the twist angle, are depicted in Fig. \ref{fig3} (d). Here, $U_A$, $U_B$, and $U_C$ represent the onsite Hubbard terms, while $V^n_{AB}$ signifies the $n$-th nearest neighbor term between the AB site, and so forth. The strength of the interaction is measured against the bandwidth of the first band. Within this system, a ``magic angle" can be identified as the angle that exhibits the highest ratio of interaction to bandwidth \cite{devakul2021magic}. This system possesses a magic angle at $\theta_m=1.75^\circ$ \cite{morales2023magic}, where the behavior of moir\'e electrons is predominantly influenced by electron-electron interactions. As the system deviates from this magic angle, the single-particle hopping increases. Nonetheless, it's crucial to recognize that the electron-electron interaction remains considerably large across the angles we are examining. 

A few remarks on the differences between the interaction parameters in the TBG scenario versus the current context are as follows: In TBG, the onsite Hubbard U is roughly on par with the nearest and next-nearest neighbor density-density interactions, a characteristic attributed to the three-peak structure of the Wannier functions. In contrast, within our model, the Wannier functions A and B exhibit strong layer polarization and commendable localization. The three-peak structure is predominantly on the layer with minimal charge components. Consequently, the density-density interaction diminishes rapidly with increasing distance as we show in Fig. \ref{fig3} (d). 
\section{A minimal two-band model}
\label{twobandmodel}

Although the three bands model is a faithful description for the low energy physics of the system, it could pose significant challenge for numerical simulations. In this section, we are aiming to reduce this model to a minimal two band model with some reasonable assumptions. 

For $\theta < 2.3^\circ$, since the total Chern number of the first two bands is zero, it is straightforward to follow above procedures for just the top two bands and derive the maximally localized Wannier functions. However, for $\theta > 2.3^\circ$, the procedure above is destined to fail due to the nontrivial topology -- the resulting Wanneir function gives wrong $C_3$ eigenvalues at $\kappa$ and $\kappa^\prime$ and the $C_{2y}\mathcal{T}$ symmetry cannot be enforced. We adopt an alternative approach. Given that the topological transition occurs between the second and third bands, which are considerably distant from the fermi energy, we expect that the correlated dynamics within the first band remain largely unaffected by this topological alteration. At the point of topological transition, the states of the second and third bands interchange in proximity to the $\gamma$ points. If we manage to retrieve the original states of the second band near $\gamma$ after the critical angle, a trivial total Hilbert space is achievable, which makes symmetric localizable Wannier functions possible. In adopting this method, the $\boldsymbol{C_3}$ eigenvalues of the second band at the $\gamma$ point do not align with the continuum model. Yet, since this discrepancy pertains to only a small section of the Brillouin zone, we expect it to have a minimal impact on the correlation physics within the first band. 


This process can be seamlessly executed in WANNIER90 through the disentanglement procedure\cite{Pizzi2020}. We set the base of the disentanglement energy window just beneath the energy of the third band at the $\gamma$ points. The Wannier functions derived from this are akin to the Wannier functions A and B in the 3-band model. The profile of the Wannier functions are illustrated in Fig. \ref{fig4}. Wannier function A(B) remains centered on the XM(MX) region, and the layer polarization is retained. This resemblance is intuitive, as we've reinstated the components of the second band, meaning the Bloch states employed for their construction are consistent with those used in the three-band model.  This is further evident in the band dispersion illustrated in Fig. \ref{fig5} (c). Here, the dispersion of the second bands, derived from the Wannier function, aligns with the third band from the continuum model at the $\gamma$ points. Notably, the first band is perfectly reproduced. We have constructed the set of maximally localized Wannier functions for a two-bands model of twisted bilayer MoTe$_2$ at the angle region $1.2^\circ \sim 3.7^\circ$. Next, we will talk about the hopping and interaction parameters based on these maximally localized Wannier functions.
 
Our results show that it is a good approximation to cutoff the hopping integral at the third nearest neighbor. With the help of $C_{2y}\mathcal{T}$ symmetry, we can write down the tight binding model:
\begin{equation}
	H= \sum_{i,j,\alpha, n=\{1,3\}}t_n\hat{c}^{\dagger}_{i\alpha}\hat{c}_{j\overline{\alpha}}+t_2\sum_{\langle\langle i,j\rangle\rangle,\alpha} e^{(-1)^{\alpha}\mathrm{i}v_{ij}\phi}\hat{c}^{\dagger}_{i\alpha}\hat{c}_{j\alpha}
\end{equation}
where $\alpha=A, B$ is index of the sub-lattice, $i$, $j$ is the unit cell index, $t_n$ is the $n$-th neighbor hopping, $t_2$ and $\phi$ are the amplitude and the phase of second nearest neighbor hopping. Including the other valley, we obtain the generalized Kane-Mele model. 
The values of the hopping parameters and their relation with the twist angles are illustrated in Fig. \ref{fig5} (a) and (b). A depiction of the band structure derived from the two-orbital tight-binding model, along with its comparison to the continuum model, can be found in Fig. \ref{fig5} (c).

We project the Coulomb interaction to the two-band model. In the two-band model, the Coulomb interaction manifests into many terms. We keep the density-density interaction up to the third nearest neighbor. Intriguingly, the projected Coulomb interaction also includes a considerable ferromagnetic spin-spin interaction and a three-site correlated hopping term. The interacting part of the Hamiltonian reads, 
\begin{equation}
\begin{aligned}
    H_{int}&=U\sum_m n_{m\uparrow}n_{m\downarrow}+\sum_{m,n,\sigma,\sigma^\prime}^{3rd NN}\bigg[(V_{m n}-\frac{1}{2}J_{mn})\hat{n}_{m\sigma}\hat{n}_{n\sigma^\prime}\\
    &-2J_{mn}\hat{S}^{z}_{m}\hat{S}^z_{n}\bigg]
    +\sum_{\langle\langle m,n\rangle\rangle,n^\prime,\sigma,\sigma^\prime}T_{mn}^{n^\prime}\hat{n}_{n^\prime}\hat{a}^{\dagger}_{m\sigma}\hat{a}_{n\sigma}  
\end{aligned}
  \end{equation}
where $m,n,n^\prime$ are indices including sub-lattice and unit cell position. $U$ is the amplitude of the onsite Hubbard interaction, $V_{mn}$ denote the strength of density-density interaction between sites, $J_{mn}$ and $T_{mn}^{n^\prime}$ are the direct spin exchange interaction and correlated hopping , which have a considerable amplitude here.  The detailed derivation can be found in Appendix \ref{fig:Sm_interaction}.

\section{Mean-field phase diagram} 
\label{meanfieldtheory}
\begin{figure*}[t]
\includegraphics[width=2\columnwidth]{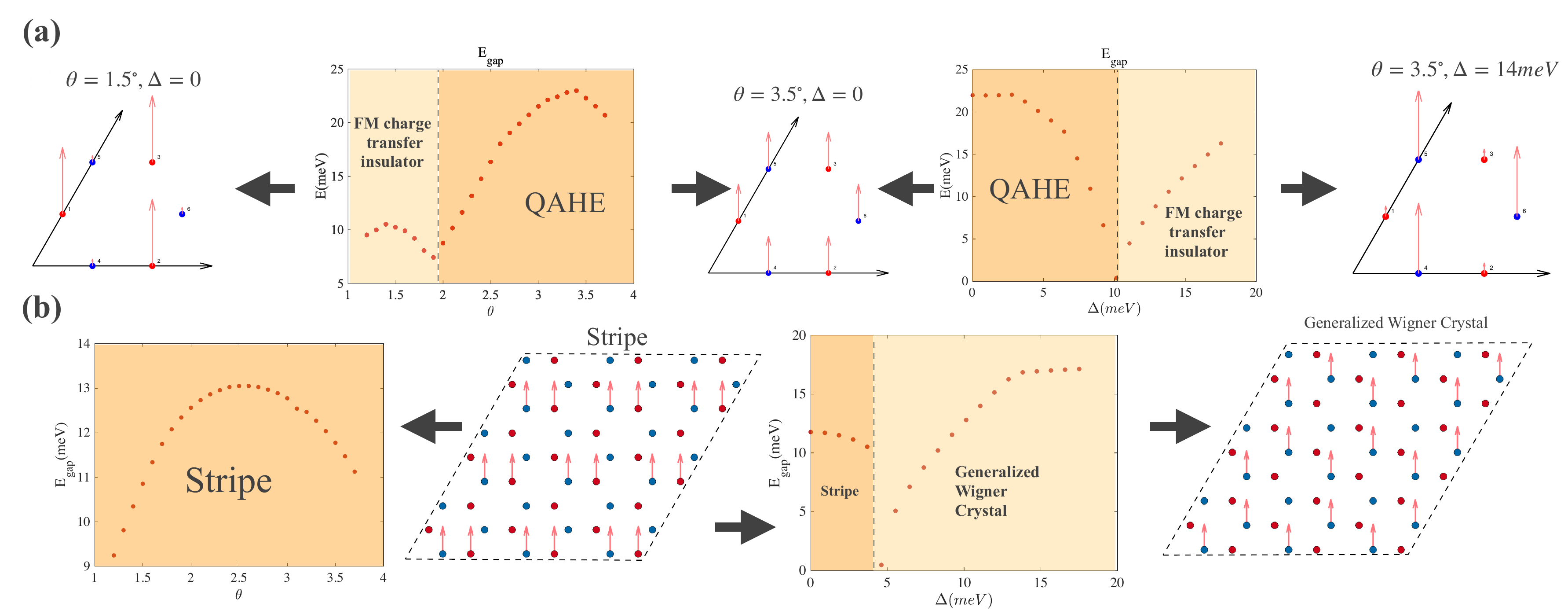}
\caption{(a) $\nu=-1$ phase diagram from Hartree-Fock stimulation. From left to right, the first panel is the charge/spin configuration of ferromagnetic charge transfer insulator. The second panel show how the gap evolves as we change the twist angles. The third panel shows the charge/spin configuration of QAH state, which is also found as the ground state with a weak gating field. And in the next panel, we show if we gradually increase the strength of gating field, there will be a continuous topological transition, which drives the system into a sub-lattice polarized charge transfer insulator.
(b) The phase diagram from Hartree-Fock stimulation at $\nu=-2/3$. In the first panel, we plot the gap as a function with the angle. The charge/spin configuration of the Stripe phase is shown in the second panel.  In the next panel, we plot the gap dependence with the gating field, which will drive the system into generalized Wigner Crystal as we show in the last panel. Here we use $\epsilon = 10$ for the dielectric constant and the screening length is $2a_m$.}
\label{fig6}
\end{figure*}

To initiate our exploration of the correlated dynamics within these moir\'e bands, we employ the Hartree-Fock mean field theory to study the minimal two-band model derived in the previous section. This allows us to examine the system's phase diagram at $\nu=-1$ and $\nu=-2/3$, considering both the twist angle \cite{qiu2023interaction} and displacement field. It's worth noting that results for $\nu=-2/3$ can only offer a preliminary insight into the correlated physics, given that the mean field theory is not adequate to capture fractional states. Nonetheless, the mean field theory is likely adept at identifying phases such as generalized Wigner Crystals at fractional fillings.

\subsection{The angle dependence at filling $\nu=-1$}
The mean field results for filling $\nu$=-1 are shown in Fig.\ref{fig6}. Here we have used a screened Coulomb potential: $V(\boldsymbol{r})=\frac{e^2}{r}e^{-r/r_0}$ with screen length $r_0=2a_m$. First, let's delve into the small angle limit, where the ground state is identified as a ferromagnetic charge transfer insulator. The substantial onsite Hubbard interaction effectively rules out double occupied states. Moreover, the pronounced nearest neighbor density-density interaction leans towards sub-lattice polarized states, which is a Mott ferroelectric insulator\cite{zhang2021electronic}. Consequently, interactions cause the charges to solidify into a triangular lattice. The remaining low-energy dynamics are governed by electron spins. At twist angle $\theta=3.5^{\circ}$, the magnitude of the antiferromagnetic exchange interaction is proportional to $\frac{t^2}{U}\sim 0.1$ meV. Our computational findings related to the ferromagnetic direct exchange coupling indicate its strength to be around $1$ meV. This suggests that a ferromagnetic phase is energetically more favorable. The synergy between the density-density interaction and direct spin exchange steers the system towards a ferromagnetic charge transfer insulator.

With an increase in the twist angle, the hopping terms also rise. This enlarged kinetic energy nudges the system towards quantum anomalous Hall (QAH) states. Observing the gap of the Hartree-Fock band structure, as depicted in Fig. \ref{fig6} (a), reveals a gap closure around the magic angle. This marks the phase boundary between topological and trivial regions. Our findings indicate that the QAH state is a spin/valley polarized state which maintains an equal charge distribution on the A and B sub-lattice.

\subsection{The gating field dependence at filling $\nu=-1$}
Recall the charge distribution of Wannier functions A and B are mostly layer-polarized. As a result, the effect of the gating field can be approximated with the onsite energy difference, namely
\begin{equation}
	H_{D}=\frac{\Delta}{2}\sum_{i\in A}\hat{c}_{i}^{\dagger}\hat{c}_{i}-\frac{\Delta}{2}\sum_{i\in B}\hat{c}_{i}^{\dagger}\hat{c}_{i}
\end{equation}
where $\Delta=\frac{Dd}{\epsilon}$, $d$ is the layer distance and $D$ is the displacement field. In scenarios with a dominant gating field, the charge is pushed into just one of the two layers, forming a triangular lattice. Given the localized character of the Wannier orbits, this state manifests as a trivial insulator. With no gating field, a QAH state is anticipated, characterized by an even charge distribution across the A and B sub-lattices. It is logical to foresee a topological phase transition as the gating field incrementally increases. Our computational findings, as illustrated in Fig. \ref{fig6} (a), indeed validate this scenario. A gap-closing transition is evident at $\Delta_c\cong10 meV$. For values of $\Delta$ less than $\Delta_c$, the system exhibits mild layer polarization but retains its QAH state. Once $\Delta$ exceeds $\Delta_c$, the ground state of the system becomes topologically trivial, continuously connected to the triangular charge transfer insulator in the large field limit. Recent experimental data has indeed observed a topological phase transition at $\nu=-1$ driven by the gating field. Our finding is consistent with this experimental result. 

\subsection{The angle dependence at $\nu=-2/3$}

The mean-field phase diagram at $\nu =-2/3$ is presented in Fig. \ref{fig6} (b). Based on our numerical findings, we observe that the ground state is a Stripe phase at the mean-field level, primarily due to the strong interaction, as we show in \ref{fig5} (d), we find the stripe phase at all twist angle. We reiterate that these findings are based on the ordinary Hartree-Fock mean-field theory which cannot encompass fractional quantum anomalous Hall state. Nonetheless, the outcomes for the strong interaction limit, namely the Stripe phase, are likely valid on general grounds.

\subsection{The gating field dependence at $\nu$=-2/3}
In this section, we focus on the system's response under a gating field at $\nu=-2/3$. As depicted in Fig. \ref{fig6} (b), a field-induced transition from the Stripe phase to generalized Wigner Crystal is evident at the mean-field level, a sub-lattice balanced states is driven into a sub-lattice polarized state. The underlying physics is similar to that observed in the integer case.

\section{Exact Diagonalization at $\nu=-2/3$} 
\label{ED}
To delve into the strongly correlated physics at this filling, we utilize band-projected momentum space exact diagonalization. A straightforward approach, adopted in previous studies\cite{morales2023pressure,PhysRevB.108.085117,wang2023fractional,dong2023composite}, involves conducting the momentum space ED using only the topmost moir\'e band. In these examinations, fractional states are typically most evident at twist angles with the narrowest bandwidth, often aligning with optimal conditions for the trace condition to be met. The typical magnitude of the gap is in the ballpark of a few meV. 

\begin{figure}[ht]
\includegraphics[width=\columnwidth]{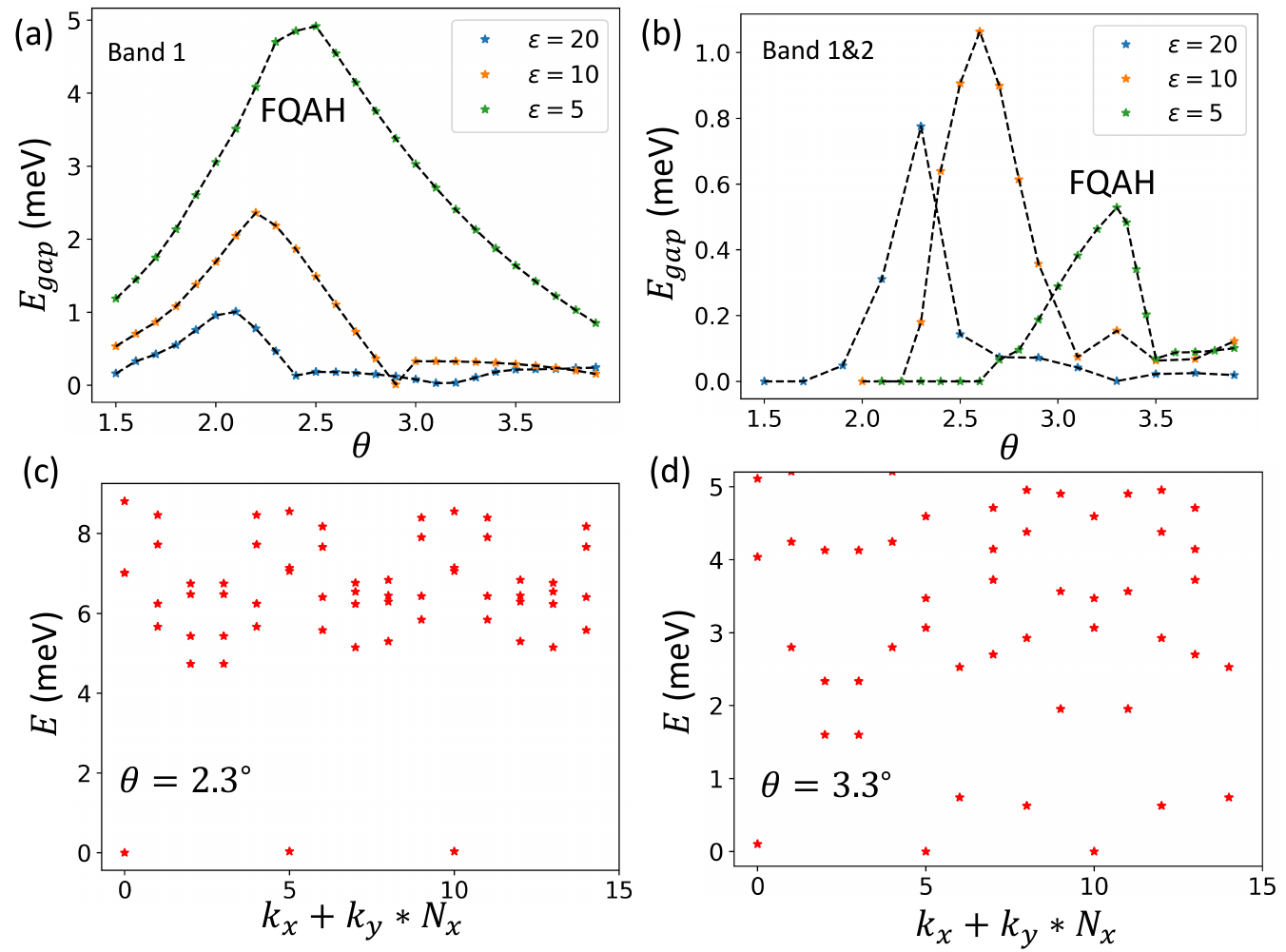}
\caption{$\nu=-\frac{2}{3}$ phase diagram from ED, the FQAH gap $E_{gap}$ is defined as the energy gap between three nearly degenerate $\nu=-2/3$ FQAH state to higher spectrum as $E_{gap}=E_4-E_3$. Here $E_i$ is the $i_{th}$ lowest energy in ED spectrum, and we set ground state energy $E_1=0$ meV.
(a) Angle dependent $\nu=-2/3$ spectrum gap from single-band projected ED, for $\epsilon=5,10,20$.
(b) Angle dependent $\nu=-2/3$ spectrum gap from two-band projected ED, for $\epsilon=5,10,20$.
(c) Single-band ED spectrum at $\theta=2.3^{\circ}$ and $\epsilon=5$.
(d) Two-band ED spectrum at $\theta=3.3^{\circ}$ and $\epsilon=5$.
}\label{fig:ED}
\end{figure}

A crucial observation we wish to highlight in this section is that the strength of the density-density interaction often surpasses the energy separation between the first and second moir\'e bands as we show  in \ref{fig:Sm_w2}. This suggests that multi-band effects, reminiscent of Landau level mixing in quantum Hall physics \cite{sodemann2013landau}, might play a significant role in this system. In this section, we will showcase a comparative analysis between the results of a one-band momentum space ED and a two-band momentum space ED. Our ED results are shown in Fig.\ref{fig:ED}, where the ground state degeneracy(threefold) matches the expectation of a fractional quantum Hall state on a torus. And the crystal momenta of the ground state manifold respect the “generalized Pauli principle” rules(SM \ref{fig:totalm}) for FQAH states \cite{PhysRevX.1.021014}

This comparison underscores notable influences of the second band. We typically find that the most pronounced FQAH state deviates from the small twist angles with flattest band, gravitating towards larger twist angles. Furthermore, the typical gap size for the FQAH state in the two-band ED is notably reduced compared to the one-band ED, often falling around 1 meV.

For the comparison to two-band projected ED, we first execute a single band-projected ED using a $k$-space lattice of dimensions $5\times 3$ and an unscreened Coulomb interaction. 
Given the strong ferromagnetism\cite{PhysRevB.108.085117}, a representative many-body spectrum in the fully polarized sector for the FQAH phase can be seen in Fig. \ref{fig:ED} (c). The gap of the FQAH state (specifically, the energy difference between three nearly degenerate states and fourth states in the many-body spectrum) as function of twist angle is illustrated in Fig. \ref{fig:ED}(a). By employing various dielectric constants, we modulate the interaction strength. The best conditions for FQAH, under a relatively mild coupling determined by a dielectric constant of $\epsilon=20$, align closely with the magic angle $\theta_m$. The maximal gap size is $1$ meV and the FQAH state only extends to $\theta\cong 2.4^\circ$. When the interaction is increased, with $\epsilon=5$ as in the experimental condition, the peak FQAH gap expands to approximately $5$ meV, and the ideal angle shifts to around $\theta\cong2.5^{\circ}$. The FQAH state persists up to $\theta\cong 4.0^{\circ}$. For larger angles, the system is expected to enter an ordinary metallic state \cite{PhysRevB.108.085117}. 



To properly take into account of the band-mixing effects, we further perform two-band projected ED as follows. Here we project the Coulomb interaction to top two moir\'e band as
\begin{equation}
\begin{aligned}
&V^{m,n,m^\prime,n^\prime}_{\boldsymbol{k_1,k^\prime_1,k_2,k_2^\prime},\sigma,\sigma^\prime}=\int d\boldsymbol{r} \int d\boldsymbol{r^\prime} \psi^*_{n\boldsymbol{k_1}\sigma}(\boldsymbol{r}) \psi_{n^\prime \boldsymbol{k_1}^\prime\sigma}(\boldsymbol{r})\\
&\frac{e^{|r-r^\prime|/r_0}}{\epsilon|r-r^\prime|}\psi_{m\boldsymbol{k_2}\sigma^\prime}(\boldsymbol{r^\prime})
	\psi^*_{m^\prime \boldsymbol{k^\prime_2}\sigma^\prime}(\boldsymbol{r^\prime}),
\end{aligned}	
\end{equation}
where $m,n,m',n'=1,2$ are the band indices, and $\sigma,\sigma'$ are the spin indices. $r_0$ is the screening length, and $\psi$'s are the Bloch wavefunctions. 

With the unscreened Coulomb interaction, we perform an ED calculation in the fully spin polarized sub-Hilbert with a $5\times3$ lattice in $k$-space. The primary characteristics of the two-band projected ED are: 1) The favored angles for the emergence of FQAH states typically shift towards larger angles, and 2) The gap sizes of the FQAH states are generally smaller than those observed in single band ED results. These observations could elucidate why experimental results more readily identify FQAH states in samples with larger twist angles ($3.3^\circ-3.7^\circ$) and not in those with smaller angles ($\sim 2^\circ$) \cite{PhysRevB.108.085117,morales2023pressure}.

The shift of the FQAH state towards larger angles can be intuitively grasped as follows: In an ideal scenario, a Chern band would be isolated in energy and meet the uniform Berry curvature and trace condition criteria. This would likely make the FQAH the dominant ground state in the presence of strong interactions. However, when other energy-proximate bands are factored in (or equivalently when the interaction strength exceeds the energy separation between bands), the system becomes susceptible to a broader array of competing states.  For example, our mean-field analysis suggests the feasibility of a Stripe phase in the strong coupling limit for the two-band model. The presence of these competing states can undermine the stability of the FQAH, making it more elusive. (It is worth noting that our current $5\times3$ lattice configuration does not align with the Stripe phase with $\sqrt{3}\times\sqrt{3}$ charge order, hindering precise phase boundary determination.) This study underscores that for FQAH states in Chern bands, relying solely on the Berry curvature and trace condition is insufficient. It's imperative to account for multi-band effects in the strong coupling regime, as these can substantially alter the underlying physics.

\section{Conclusion and outlook }

In this paper, we delve deeply into the single-particle and many-body physics of the twisted homobilayer MoTe$_2$ system. Our work begins with a comprehensive exploration of the moir\'e band structure, where we harness the power of large-scale density functional theory calculations to fit accurate parameters for the continuum model. Our calculations reveal that, at experimentally pertinent twist angles, the top three moir\'e bands consistently display nontrivial Chern numbers, but together they are topologically trivial.

To set the stage for studying the correlated physics within these bands, we extracted the maximally localized Wannier functions for the top three moir\'e bands. By making certain assumptions, we condensed our model into a minimal two-band version via band disentanglement. A standout feature of this model is the pronounced layer polarization exhibited by the Wannier orbitals located at the XM and MX regions. This refined model serves as a foundation for future investigations into the topological and correlated physics in twisted MoTe$_2$ system.

We performed a mean field calculation to map out the phase diagram of the two-band model at fillings of $\nu=-1$ and $\nu=-2/3$, analyzed as functions of twist angles and displacement fields. Specifically, at a filling of $\nu=-1$, we identified a field-driven continuous topological transition from the quantum anomalous Hall state to a trivial charge transfer insulator, consistent with experimental findings \cite{park2023observation,xu2023observation}.

Finally, we employed band-projected momentum space exact diagonalization to explore the FQAH states in this system. As our calculation shows that the interaction strength in this system easily excesses the energy separation between the first and second moir\'e bands, we anticipate strong band-mixing effects in this system. Indeed, a comparative study between one-band ED and two-band ED unveiled stark contrasts. A key observation was the shift in the optimal twist angle for the FQAH state in the two-band system, moving away from the magic angle where the first band has the smallest bandwidth. This shift is likely attributed to the intense competition with other potential orderings in the multi-band Hilbert space at these strongly interacting situations. Such findings underscore the need for a careful examination of multi-band physics in this system in upcoming studies.

Given the limitations of current computational capabilities, an ED analysis of the minimal two-band or the three-band model in real space remains challenging given the $2a_m$ cutoff at large twist angles. However, as we have emphasized, understanding the multi-band physics is pivotal for comprehending the stability of the FQAH state, especially in the context of the FQAH regime with a large twist angle. It becomes imperative to venture into other numerical methods such as Density Matrix Renormalization Group (DMRG) simulations. This exploration will be left for future works.

In experimental observations \cite{park2023observation,xu2023observation}, a captivating phase transition emerges at the fractional filling $\nu=-2/3$, where the FQAH state transitions to a more conventional insulating state under the influence of a gating field. Our preliminary ED studies have observed a gap closure of the FQAH state when a gating field is applied. Such a transition challenges the traditional Landau-Ginzburg paradigm, suggesting the potential presence of exotic quantum critical points in the phase diagram. The nature of this gate-induced phase transition remains an exciting direction for future research. Moreover, a compressible state at $\nu=-1/2$ has been identified \cite{xu2023observation}. Both experimental and theoretical perspectives \cite{dong2023composite,goldman2023zero} have yet to fully elucidate the nature of this state. We anticipate that our model will serve as a foundational platform to investigate these intriguing intriguing compressible states and global phase diagram further \cite{reddy2023toward}. Ideally, this exploration will also illuminate the optimal conditions for realizing even-denominator fractional quantum anomalous Hall states in the TMD moir\'e systems.

\section*{Acknowledgments}
We are grateful to Kin Fai Mak, Yihang Zeng, Liang Fu, Jie Wang, Jainendra Jain, Tian-Sheng Zeng, and Cristian D. Batista for helpful discussions. Y. Z. thanks Tingxin Li for the experimental collaboration on transport probe of FQAH in twisted MoTe$_2$. Z. B. is supported by the start-up fund at the Pennsylvania State University. Y. Z. is supported by the start-up fund at University of Tennessee Knoxville.

\bibliographystyle{apsrev4-2}
\bibliography{moire_MoTe.bib}
\clearpage
\pagebreak
\onecolumngrid
\begin{center}
\textbf{\large Appendix}
\end{center}
\setcounter{figure}{0}
\renewcommand{\thefigure}{S\arabic{figure}}
\setcounter{equation}{0}
\renewcommand{\theequation}{S\arabic{equation}}
\setcounter{table}{0}
\renewcommand{\thetable}{S\arabic{table}}

\subsection{Density functional theory calculation and continuum model}
We study TMD homobilayers with a small twist angle starting from AA stacking, where every metal (M) or chalcogen (X) atom  on  the top layer is aligned with the same type of atom on the bottom layer. 
Within a local region of a twisted bilayer, the atom configuration is identical to that of an untwisted bilayer, where one layer is laterally shifted relative to the other layer by a corresponding displacement vector ${\bm d}_0$. For this reason, the moir\'e band structures of twisted TMD bilayers can be constructed from a family of untwisted bilayers at various ${\bm d}_0$, all having $1\times 1$ unit cell. Our analysis thus starts from untwisted bilayers.

In particular, ${\bm d}_0=0, -\left({\bm a}_{1}+{\bm a}_{2}\right) /3, \left({\bm a}_{1}+{\bm a}_{2}\right) /3$, where ${\bm a}_{1,2}$ is the primitive lattice vector  for untwisted bilayers, correspond to three high-symmetry
stacking configurations of untwisted TMD bilayers, which we refer to as MM, XM, MX. In MM (MX) stacking, the M atom on the top layer is locally aligned with the M (X) atom on the bottom layer, likewise for XM. The bilayer structure in these stacking configurations is invariant under three-fold rotation around the $z$ axis.

\subsubsection{Testing of different vdW corrections}

\renewcommand{\arraystretch}{1.15}
\renewcommand{\multirowsetup}{\centering}

It is well known that different vdW corrections lead to different relaxed layer distances in TMD materials, here we performed the testing with ten different vdW functionals as implemented in VASP. We first construct the zero-twisted angle MoTe$_2$ bilayer at MM and MX lateral configurations with vacuum spacing larger than 20\AA{} to avoid artificial interaction between the periodic images along the z direction. Four different in-plane lattice constants have been tested 3.52\AA{} (from bulk structures), 3.472\AA{} (relaxed lattice constant with PBE), 3.49\AA{} (low temperature considering thermal expansion), and 3.57\AA{} (from 2D material database). The structure relaxation is performed with force on each atom less than 0.001 eV/A. We use $12\times 12 \times 1$ for structure relaxation and self-consistent calculation. 

We note SCAN+rVV10, IVDW 4, 10, 11, 21 dispersion correction method gives the relaxed layer distances around 7.0\AA{} (close to the bulk layer distance 6.98\AA{} in 2H structures) and around 7.65 t0 7.80\AA{} for MX and MM stacking structures, respectively. By calculating the work function from electrostatic energy of converged charge density, we can determine energy difference of valence band maximum at $\Gamma$ and $K$ valley in MM and MX-stacked bilayers, with reference energy chosen to be the absolute vacuum level. Since the dipole correction and specific vdW functional (different layer distance) heavily affected the numerical stability, there will be a large fluctuation in continuum model fitting with the shift methods, which we will study in details in a separate work.

\subsubsection{Large-scale density functional theory calculations and continuum model parameters}
To obtain the accurate continuum model parameter, we perform large-scale density functional theory calculations with various van der Walls functionals. From large-scale DFT at $\theta=4.4^\circ$ with 1014 atoms per unit cell in previous work, we obtain the continuum model parameters $V=11.2$ meV, $\phi=-91^\circ$ and $w=-13.3$meV, with the bulk lattice constant $a_0 = 3.52\text{\AA{}}$ and the effective mass $m^* = 0.62 m_e$. 

At twist angle $\theta=3.89^\circ$, we relax the lattice structures with computationally low-cost vdW functionals, and fit the continuum model parameters as: (a) IVDW 4, $m^*=0.62m_e,V=9.2  \ meV, w=-11.2\  meV,\phi=-99^\circ$. 
(b) IVDW21, $m^*=0.62m_e, V=7.5$meV, $\psi=-100^\circ$ and $w=-11.3$meV. In this work, we choose the parameter from IVDW 4, which best reproduces the bulk layer distance in 2H structure. 

The Chern number is calculated from symmetry eigenvalues at three high symmetry points:
\begin{equation}
e^{i 2 \pi C / 3}=\prod_{i \in \text { occ. }}(-1)^F \theta_i(\Gamma) \theta_i(K) \theta_i\left(K^{\prime}\right)
\end{equation}
Where $F$ is two times the total spin
of the particle, and 1 for the spin $\frac{1}{2}$ in the valence moir\'e bands. And $\theta_i(\mathbf{k})$ represents the $C_3$ symmetry eigenvalues
at $C_3$-invariant k points. Computationally, the Kohn-Sham wavefunctions in plane-wave basis $$
\left|\Psi_{n \boldsymbol{k}}\right\rangle=\sum_{\boldsymbol{G}} C_{n \boldsymbol{k}}(\mathbf{G})|\boldsymbol{k}+\boldsymbol{G}\rangle
$$
are extracted at three $C_{3z}$ invariant momentum $\gamma$, $\kappa$, and $\kappa'$ with energy cutoff as $E_{cut}=100$ eV and $\hbar^{2}(\boldsymbol{k}+\boldsymbol{G})^{2} / 2 m_{e}<E_{\mathrm{cut}} $. And we compute their $C_{3z}$ eigenvalue as:
\begin{equation}
\left\langle\Psi_{n k}|C_{3z}| \Psi_{n k}\right\rangle=\sum_{\tau \tau^{\prime}} S_{\sigma \sigma^{\prime}}(C_{3z})\left\langle\Psi_{n k}^{\sigma}|C_{3z}| \Psi_{n k}^{\sigma^{\prime}}\right\rangle
\end{equation}
where $\sigma,\sigma'=\uparrow$ or $\downarrow$, $\left\langle\Psi_{n k}^{\sigma}|C_{3z}| \Psi_{n k}^{\sigma^{\prime}}\right\rangle$ calculated from the transformation relation of plane-waves, and $S_{\sigma \sigma^{\prime}}(C_{3z})$ is the $C_{3z}$ spin rotation matrix.

As there is a valley time reversal symmetry within the moir\'e superlattice, the energy bands between $\gamma$, $\kappa$ and $\kappa'$ are enforced to be double degenerate from $K$ and $-K$ valley. Therefore, to determine the valley index of moir\'e bands, we add a small gating field $D=0.005$ V/nm to introduce a less than 2 meV splitting at $\kappa$ and $\kappa'$. The $K$ and $-K$ indexes are then diagnosed by spin resolved wavefunction.
At $\gamma$ point where bands are still double degenerate with gating field, we extracted the $C_{3z}$ eigenvalues and valley indexes from finite momentum perturbation along $\gamma$ to $\kappa$ direction. 

For homobilayer MoTe$_2$ at $\theta=4.4^{\circ}$, we have $e^{-i\pi/3}$, $e^{i\pi/3}$, and $e^{i\pi/3}$ for $\gamma$, $\kappa$ and $\kappa'$, respectively. Therefore the corresponding Chern number is $C=-1$.




\subsubsection{Topological edge states at $\nu=-2$ and $\nu=-4$}
\begin{figure}[hbt]  \label{EdgeStates}
\includegraphics[width=0.8\textwidth]{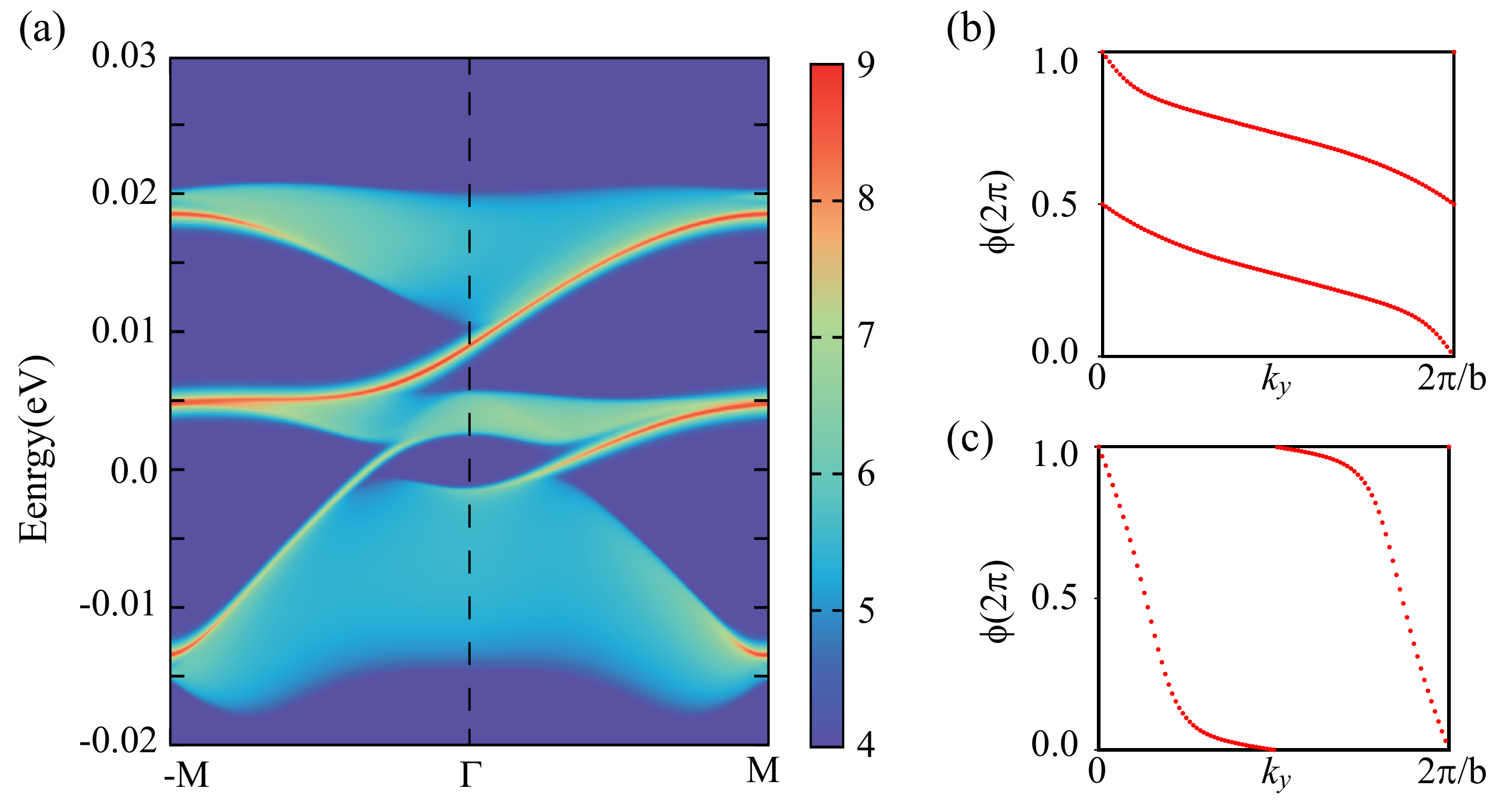}
  \caption{(a) Topological edge states along the high-symmetry line.
  The Wannier charge centers along $k_y$ for the lowest conduction band (b) and highest valence band (c). }
  \label{wilsonline}
\end{figure}

For the experimental twist angle $\theta=3.5^{\circ}$, the top three moir\'e valence bands are all topological, with Chern number -1, -1, and 2 for $K$ valley. Therefore at single particle level, there exist a quantum spin Hall edge state at $\nu=-2$ and a double quantum spin Hall edge states at at $\nu=-4$ as shown in Fig. \ref{wilsonline}.

\subsection{The Coulomb interaction in full polarized sub-space}
\begin{figure}[hbt]
  \includegraphics[width=1\textwidth]{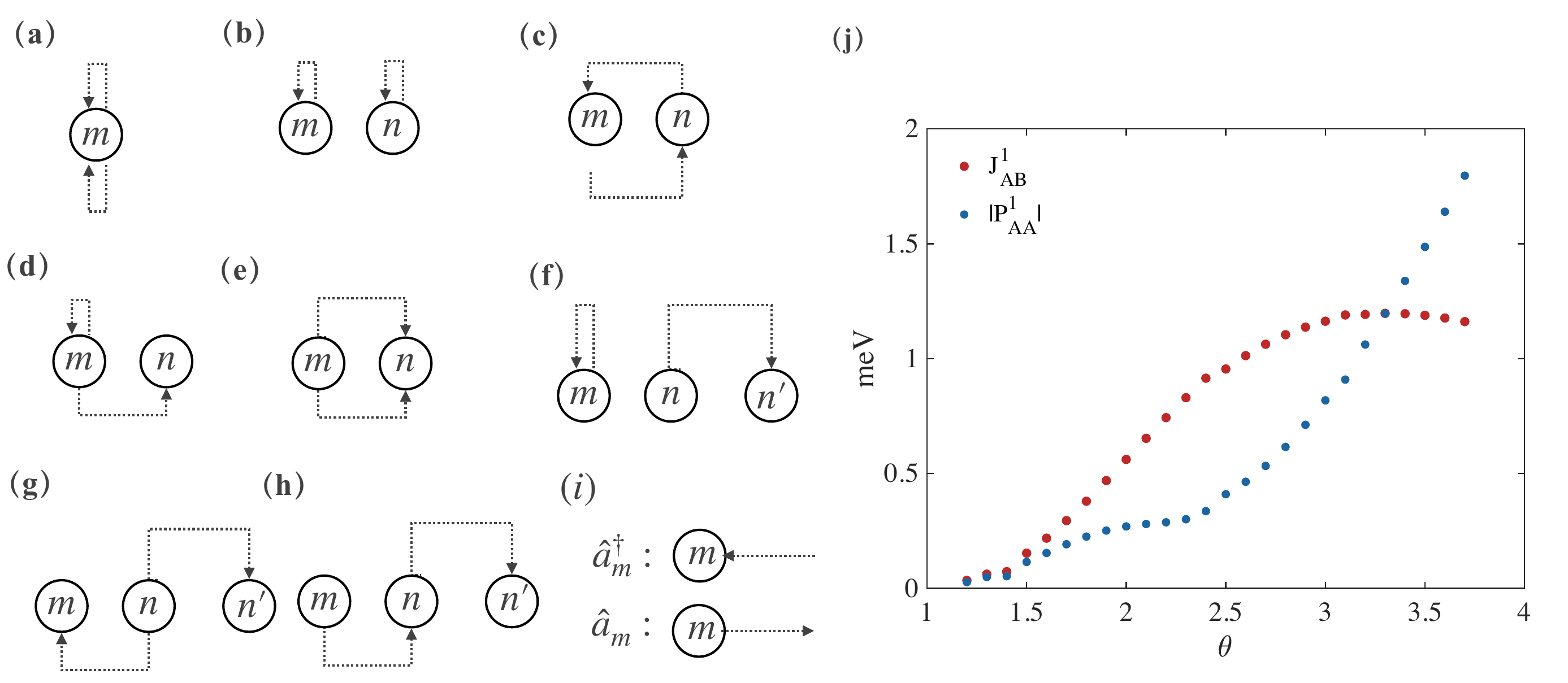}
  \caption{(a)-(f) The graphic representation of the four fermion operators. The creation operator is expressed with an arrow ending at a site, while an arrow going away from the site represents annihilating a particle at that site.
  (a) the Hubbard interaction; 
  (b) the density-density interaction; 
  (c) the direct exchange interaction; 
  (d) the onsite correlated hopping interaction;
  (e) the local pair hopping; 
  (f) the correlated hopping; 
  (g) the off-site local pair hopping 
  (h) the indirect correlated hopping; 
  (j) the amplitude of the direct spin exchange and correlated hopping as a function of twist angle, the screening length is chosen to be $r_0=2a_m$ here and $\epsilon =10$.} 
  \label{fig:Sm_interaction}
\end{figure}

The general expression of  Coulomb interaction reads:
\begin{equation}
	\begin{aligned}
		V=\frac{1}{2}\int\int d\boldsymbol{r}d\boldsymbol{r}^\prime \frac{e^{-|\boldsymbol{r-r^\prime}|/r_0}}{\epsilon|\boldsymbol{r-r^\prime}|}\hat{c}_{\sigma}^{\dagger}(\boldsymbol{r})\hat{c}_{\sigma^\prime}^{\dagger}(\boldsymbol{r^\prime})c_{\sigma^\prime}(\boldsymbol{r^\prime})c_{\sigma}(\boldsymbol{r})
	\end{aligned}
\end{equation}
We project the Coulomb interaction to the Wannier function basis:
\begin{equation}
	c^{\dagger}_{\sigma}(\boldsymbol{r})=\sum_{m}w_{m}^{*}(\boldsymbol{r})a^{\dagger}_{m\sigma}
\end{equation}
where m is the indice including sub-lattice and unit cell position. The strength of the general interaction reads:

\begin{equation}
	\begin{aligned}
		V&=\frac{1}{2}\sum_{m,n,m^\prime,n^\prime}\sum_{\sigma\sigma^\prime}\bigg(\int\int d\boldsymbol{r}d\boldsymbol{r^\prime}w_{m}^*(\boldsymbol{r})w_{n}^*(\boldsymbol{r^\prime}) \frac{e^{-|\boldsymbol{r-r^\prime}|/r_0}}{\epsilon|\boldsymbol{r-r^\prime}|}w_{n^\prime}(\boldsymbol{r})w_{m^\prime}(\boldsymbol{r^\prime})\bigg)a^{\dagger}_{m\sigma}a^{\dagger}_{n\sigma^\prime}a_{n^\prime\sigma^\prime}a_{m^\prime\sigma}\\
		&=\sum_{m,n,m^\prime,n^\prime}\sum_{\sigma\sigma^\prime}V_{m^\prime n^\prime}^{mn}a^{\dagger}_{m\sigma}a^{\dagger}_{n\sigma^\prime}a_{n^\prime\sigma^\prime}a_{m^\prime\sigma}\\
	\end{aligned}
\end{equation}
With this equation, the hopping process of  electrons journeying through the lattice become interconnected due to the influence of Coulomb interaction. These components can be classified based on the quantity of sites they encompass. We show the graphic representation of interaction up to three site term in the Fig \ref{fig:Sm_interaction}. In our main text, we keep these term:
\begin{equation}
\begin{aligned}
    H_{int}&=U\sum_m n_{m\uparrow}n_{m\downarrow}+\sum_{m,n,\sigma,\sigma^\prime}^{3rd NN}\bigg[(V_{m n}-\frac{1}{2}J_{mn})\hat{n}_{m\sigma}\hat{n}_{n\sigma^\prime}\\
    &-2J_{mn}\hat{S}^{z}_{m}\hat{S}^z_{n}\bigg]
    +\sum_{\langle\langle m,n\rangle\rangle,n^\prime,\sigma,\sigma^\prime}T_{mn}^{n^\prime}\hat{n}_{n^\prime}\hat{a}^{\dagger}_{m\sigma}\hat{a}_{n\sigma}  
\end{aligned}
  \end{equation}
We further show how the direct exchange term and correlated hopping term change with angle in Fig\ref{fig:Sm_interaction}. And due to the valley polarized state is preferable, we focus on the interaction in the fully polarized state in this part, which means the doubly occupied site is forbidden owing to the Pauli exclusion principle. With our expression in Fig\ref{fig:Sm_interaction}, the site that is the starting points(or endpoints) of 2 arrow is forbidden. The left terms are  density-density interaction(Fig\ref{fig:Sm_interaction}.(b)), direct exchange interaction (Fig\ref{fig:Sm_interaction}.(c)), correlated hopping (Fig\ref{fig:Sm_interaction}.(f)). So, we finally arrive the interaction model in the fully polarized sub-space:

\begin{equation}
    H_{int}=\sum_{m,n}^{3rdNN}\bigg[(V_{m n}-\frac{1}{2}J_{mn})\hat{n}_{m}\hat{n}_{n}-2J_{mn}\hat{S}^{z}_{m}\hat{S}^z_{n}\bigg]+\sum_{\langle\langle m,n\rangle\rangle,n^\prime,\sigma,\sigma^\prime}T_{mn}^{n^\prime}\hat{n}_{n^\prime}\hat{a}^{\dagger}_{m\sigma}\hat{a}_{n\sigma} 
    \end{equation}

\subsection{Construct the Symmetry adapted Wannier function}
\subsubsection{Basis of the continuum model}
Here we present a detailed description of the procedure we use to construct the Wannier function in the main text. Before entering the main part, I would like to discuss the basis we use in the continuum model. The low energy electron we include in the continuum model is the Bloch states of the single layer TMD near the $\boldsymbol{K_{\pm}}$, the interlayer and intra-layer coupling will couple them and give the final moire bands. We label the momentum of the Bloch states as:
\begin{equation}
	\begin{aligned}
		\boldsymbol{k_t}&=\boldsymbol{K_{+}-q_3+k_{t}^{m}}\\
				\boldsymbol{k_b}&=\boldsymbol{K_{-}+q_2+k_{b}^{m}}
	\end{aligned}
\end{equation}
where $k_{t}$ is the Bloch states in the top layer and it is measured relative to the $\Gamma$ points of the single layer TMD and $k_{t}^{m}$ is measured relative to the $\Gamma_{M}$ of the moire Brillouin zone(mBZ), $\boldsymbol{q_1}$ is the momentum difference between the $K_{+}$ and $K_{-}$, $\boldsymbol{q_2}(\bm{q_3})$ is the $C_3(C_3^{-1})$ operate on the $\bm{q_1}$. We further restrict the momentum in the 1st mBZ and the formula will be:
\begin{equation}
	\begin{aligned}
		\boldsymbol{k_t}&=\boldsymbol{K_{+}-q_3+k_{t}^{m}+G_m}\\
				\boldsymbol{k_b}&=\boldsymbol{K_{-}+q_2+k_{b}^{m}+G_m}
	\end{aligned}
\end{equation}
\begin{figure}[hbt]
  \includegraphics[width=1\textwidth]{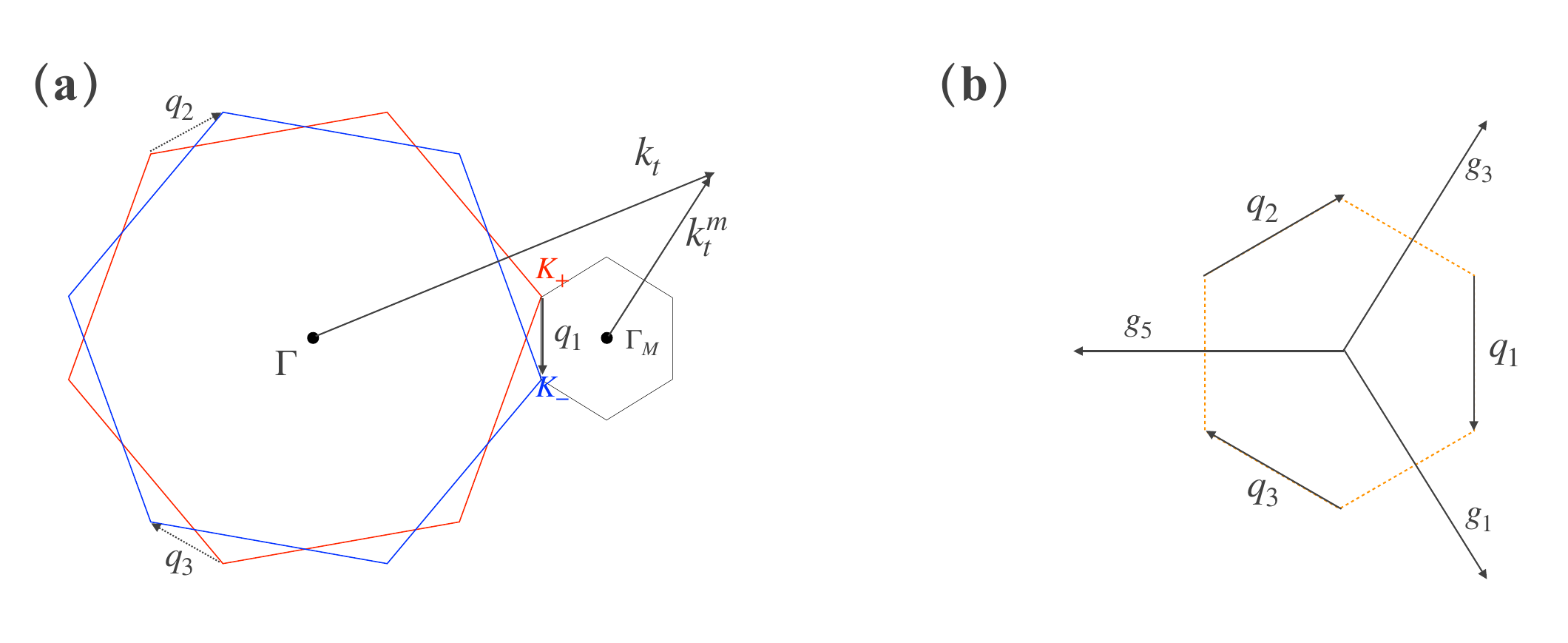}
  \caption{(a) The relation between the Brillouin zone(BZ) of single layer MoTe$_2$ and the moir\'e Brolliun zone(mBZ). The red(blue) hexagon is the BZ of top(bottom) layer MoTe$_2$ and the mBZ is shown with the small black hexagon. $\Gamma,K_{\pm}$ is the high symmetry points of single layer MoTe$_2$, $q_1,q_2,q_3$ are the momentum difference between two layer. $k_t$($k_t^m$) is the momentum of the plane wave basis which is measured realtive to the original point of BZ(mBZ). 
  (b) the relation between the $q_1,q_2,q_3$ and mBZ, where $g_1,g_3$ are the reciprocal vector of mBZ.}\label{fig:Sm_mBZ}
\end{figure}
Note that $k_{t}^{m}$ and $k_{b}^{m}$ are in the 1st mBZ here. Then we use the relation between the $q_i$ and moire reciprocal vector, we finally arrive:
\begin{equation}
	\begin{aligned}
		\boldsymbol{k_t}&=\boldsymbol{K_{+}+k-Q_{+}}\\
				\boldsymbol{k_b}&=\boldsymbol{K_{-}+k-Q_{-}}
	\end{aligned}
\end{equation}
where $\boldsymbol{Q_{\pm}=\pm q_1+G_{m}}$, they form  a honeycomb lattice in K-space and all of symmetry can be kept with any finite cutoff, which gives a very convenient representation of the continuum model. The inter-layer interaction will transform into the coupling between the nearest neighbor momentum in the lattice and the intra-layer coupling will be the nearest intra-sub-lattice coupling. And then we introduce a continuous real space basis:
\begin{equation}
	|l,\boldsymbol{r}\rangle =\frac{1}{\sqrt{\Omega_{tot}}}\sum_{\boldsymbol{k}}\sum_{\boldsymbol{Q}\in \boldsymbol{Q_{l}}} e^{-i(\boldsymbol{k-Q})\cdot\boldsymbol{r}}|\boldsymbol{k,Q}\rangle
\end{equation}
where $\Omega_{tot}$ is the total area of the system, $|\boldsymbol{k,Q}>$ is the Bloch state of single layer TMD. The we can write down the expression of the continuum model in the Bloch states of single layer TMD:
\begin{equation}
	\begin{aligned}
		H&=-\sum_{\boldsymbol{Q\in Q_0}}\frac{(\boldsymbol{k-Q})^2}{2m^*}+\sum_{\boldsymbol{Q,Q^\prime}\in \boldsymbol{Q_0},j=1}^{3}w\delta_{\boldsymbol{Q,Q^\prime\pm q_j}}\\
  &+\sum_{l=\pm}\sum_{\boldsymbol{Q,Q^\prime\in Q_l},i=1,3,5}V(e^{\mathrm{i}l\phi}\delta_{\boldsymbol{Q,Q^\prime+g_i}}+e^{-\mathrm{i}l\phi}\delta_{\boldsymbol{Q,Q^\prime-g_i}})
	\end{aligned}
\end{equation}
where $l=\pm$ is the layer index. We next diagonalize the Hamiltonian to get the eigenstate:
\begin{equation}
	|\psi_{n\boldsymbol{k}}\rangle=\sum_{\bm{Q\in Q_0}}C_{n\boldsymbol{k}}(\bm{Q})|\boldsymbol{k,Q}\rangle
\end{equation}
Where the $C_{n\boldsymbol{k}}(\bm{Q})$ is the eigen-vector of the Hamiltonian. The real space distribution of the eigenstate is calculated with the real space continuous space basis:
\begin{equation}
\begin{aligned}
	\psi^{l}_{n\boldsymbol{k}}(\boldsymbol{r})=\langle l,\boldsymbol{r}|\psi_{n\boldsymbol{k}}\rangle&=\sum_{\bm{Q\in Q_0}}C_{n\boldsymbol{k}}(\bm{Q})\langle l,\boldsymbol{r}|\boldsymbol{k,Q}\rangle\\
	&=\frac{1}{\sqrt{\Omega_{tot}}}\sum_{\bm{Q\in Q_l}}C_{n\boldsymbol{k}}(\bm{Q})e^{i(\boldsymbol{k-Q})\cdot\boldsymbol{r}}
\end{aligned}
\end{equation}
From this expression, we can see what we mean by using the plane wave as the basis.

\subsubsection{Generate the input file of WANNIER90 }
In our work, we use the WANNIER90 to minimize the spread of Wannier function and it needs some input files: "seedname.win","seedname.eig","seedname.mmn", "seedname.amn","seedname.dmn". Getting the first two files is not a problem, our main concern is how to produce the last three files. Firstly, the overlap between the Bloch state(the periodic part):
\begin{equation}
	\begin{aligned}
		\int d\boldsymbol{r} u^*_{m\boldsymbol{k_1}}(\boldsymbol{r})u_{n\boldsymbol{k_2}}(\boldsymbol{r})&=\frac{1}{\Omega_{tot}}\sum_{\boldsymbol{Q,Q^\prime}}C^*_{m\boldsymbol{k_1}}(\boldsymbol{Q^\prime})C_{n\boldsymbol{k_2}}(\boldsymbol{Q})\int d\boldsymbol{r} e^{i(\boldsymbol{Q^\prime-Q})\cdot\boldsymbol{r}}\\
		&=\sum_{\boldsymbol{Q,Q^\prime}}C^*_{m\boldsymbol{k_1}}(\boldsymbol{Q^\prime})C_{n\boldsymbol{k_2}}(\boldsymbol{Q})\delta_{\boldsymbol{Q,Q^\prime}}\\
		&=\sum_{\boldsymbol{Q}}C^*_{m\boldsymbol{k_1}}(\boldsymbol{Q})C_{n\boldsymbol{k_2}}(\boldsymbol{Q})
	\end{aligned}
\end{equation} 
where $u_{n\boldsymbol{k}}(\boldsymbol{r})=e^{-i\boldsymbol{k\cdot r}}\psi_{n\boldsymbol{k}}(\boldsymbol{r})$ is the periodic part of the Bloch states. It is easy to find that the quantity is just the matrix times between the eigenvector from the diagonalization of Hamiltonian. And this is the information that the "seedname.mmn" contain, the left thing to do is transforming it into the format that WANNIER90 want, one can know how to do it from the handbook of the WANNIER90.  In principle, the overlap matrix is the only thing we need to calculate and minimize the spread of Wannier function. However, a good initial guess always help us a lot, especially in the case of entangled bands. The trial Wannier function is written into the file: "seedname.amn". One can calculate it directly:
\begin{equation}
\begin{aligned}
	A_{mn}(\boldsymbol{k})&=\int \psi^*_{nk}(\boldsymbol{r})g_m(\boldsymbol{r})d\boldsymbol{r}
	=\frac{1}{\sqrt{\Omega_{tot}}}\sum_{\bm{Q\in Q}}C^*_{n\boldsymbol{k}}(\bm{Q})\int d\boldsymbol{r} g_m(\boldsymbol{r})e^{-i(\boldsymbol{k-Q})\cdot\boldsymbol{r}}
\end{aligned}	
\end{equation}
where $g_{m}(\boldsymbol{r})$ is the trial Wannier function we choose. Now let's move forward to the symmetry related file: "seedname.dmn". We choose  $C_3$ symmetry as an example to show how to calculate it:
\begin{equation}
	\begin{aligned}
		d_{mn}(C_3,\boldsymbol{k})&=\int d\boldsymbol{r} \psi^*_{m\boldsymbol{C_3k}}(\boldsymbol{r})\boldsymbol{\hat{C}_3}\psi_{n\boldsymbol{k}}(\boldsymbol{r})\\
		&=\frac{1}{\Omega_{tot}}\sum_{\boldsymbol{Q,Q^\prime}}C^{*}_{m\boldsymbol{C_3k}}(Q^\prime)C_{n\boldsymbol{k}}(Q)\int d\boldsymbol{r}e^{-i(C_3\boldsymbol{k}-\boldsymbol{Q^\prime})\cdot\boldsymbol{r}}\boldsymbol{\hat{C}_3}e^{i(\boldsymbol{k-Q})\cdot \boldsymbol{r}}\\
		&=\frac{1}{\Omega_{tot}}\sum_{\boldsymbol{Q,Q^\prime}}C^{*}_{m\boldsymbol{C_3k}}(Q^\prime)C_{n\boldsymbol{k}}(Q)\int d\boldsymbol{r}e^{-i(C_3\boldsymbol{k}-\boldsymbol{Q^\prime})\cdot\boldsymbol{r}}e^{i(\boldsymbol{k-Q})\cdot C_3^{-1}\boldsymbol{r}}\\
		&=\frac{1}{\Omega_{tot}}\sum_{\boldsymbol{Q,Q^\prime}}C^{*}_{m\boldsymbol{C_3k}}(Q^\prime)C_{n\boldsymbol{k}}(Q)\int d\boldsymbol{r}e^{-i(C_3\boldsymbol{k}-\boldsymbol{Q^\prime})\cdot\boldsymbol{r}+i(C_3\boldsymbol{k}-C_3\boldsymbol{Q})\cdot\boldsymbol{r}}\\
		&=\sum_{\boldsymbol{Q,Q^\prime}}C^{*}_{m\boldsymbol{C_3k}}(Q^\prime)C_{n\boldsymbol{k}}(Q)\delta_{Q^\prime,C_3Q}\\
		&=\sum_{\boldsymbol{Q}}C^{*}_{m\boldsymbol{C_3k}}(C_3Q)C_{n\boldsymbol{k}}(Q)\label{dmn}
	\end{aligned}
\end{equation}
One should carefully deal with the embedding matrix whenever you move the eigen-state in K space with reciprocal vectors. The $d_{mn}$ tell us how the Bloch state transform under the symmetry operation.  We further choose the behavior of Wannier function under $C_3$ reads:
\begin{equation}
\begin{aligned}
	\hat{C}_3w(\boldsymbol{r-R-\tau_n})&=e^{i\phi_n}w(\boldsymbol{r-R^\prime-\tau_n})\\
C_3(\boldsymbol{	R+\tau_n})&=\boldsymbol{R^\prime+\tau_n}
\end{aligned}
\end{equation} 
where $\boldsymbol{\tau_n}$ is the position vector of the Wannier function n in the unit cell.  The symmetry of Wannier function is indicated with the Fourier transformation of the Wannier function:
 \begin{equation}
 	\begin{aligned}
 	D_{mn}(C_3,\boldsymbol{k})&=\int d\boldsymbol{r} \tilde{\psi}^*_{m\boldsymbol{C_3k}}(\boldsymbol{r})\boldsymbol{\hat{C}_3}\tilde{\psi}_{n\boldsymbol{k}}(\boldsymbol{r})\\	
 	&=\int d\boldsymbol{r} \tilde{\psi}^*_{m\boldsymbol{C_3k}}(\boldsymbol{r})\boldsymbol{\hat{C}_3}\frac{1}{\sqrt{N}}\sum_{\boldsymbol{R}}
e^{i\boldsymbol{k\cdot R}}w(\boldsymbol{r-R-\tau_n})\\
& =e^{i\phi_{n}}\int d\boldsymbol{r} \tilde{\psi}^*_{m\boldsymbol{C_3k}}(\boldsymbol{r})\frac{1}{\sqrt{N}}\sum_{\boldsymbol{R}}e^{i\boldsymbol{k\cdot R}}w(\boldsymbol{r-R^\prime-\tau_n})\\
&=e^{i\phi_{n}}\int d\boldsymbol{r} \tilde{\psi}^*_{m\boldsymbol{C_3k}}(\boldsymbol{r})\frac{1}{\sqrt{N}}\sum_{\boldsymbol{R^\prime}}e^{i\boldsymbol{k}\cdot [C_3^{-1}(\boldsymbol{R^\prime+\tau_n})-\boldsymbol{\tau_n}]}w(\boldsymbol{r-R^\prime-\tau_n})
\\
&=e^{i\phi_{n}+i(C_3\boldsymbol{k}-\boldsymbol{k})\cdot\boldsymbol{\tau_n}}\int d\boldsymbol{r} \tilde{\psi}^*_{m\boldsymbol{C_3k}}
\frac{1}{\sqrt{N}}\sum_{\boldsymbol{R^\prime}}e^{i\boldsymbol{C_3k\cdot R^\prime}}w(\boldsymbol{r-R^\prime-\tau_n})\\
&=e^{i\phi_{n}+i(C_3\boldsymbol{k}-\boldsymbol{k})\cdot\boldsymbol{\tau_n}}\int d\boldsymbol{r} \tilde{\psi}^*_{m\boldsymbol{C_3k}}
\tilde{\psi}_{n\boldsymbol{C_3k}}\\
&=e^{i\phi_{n}} e^{i(C_3\boldsymbol{k}-\boldsymbol{k})\cdot\boldsymbol{\tau_n}}\delta_{mn}
 	\end{aligned}
 \end{equation}
where $\boldsymbol{\tau_n}$ is the position vector of the Wannier function n in the unit cell and 
the site symmetry group associated with $\tau_n$ should align with the representation of the band structure.
\subsection{The interaction parameter measured with first two bands $W_2$}
In this part, we show that the strength of density-density interaction surpasses the separation between the first band and second moire bands. This suggests that multi-band effects, reminiscent of Landau level mixing in quantum Hall physics \cite{sodemann2013landau}, might play a significant role in this system.
\begin{figure}[h]
  \includegraphics[width=1\textwidth]{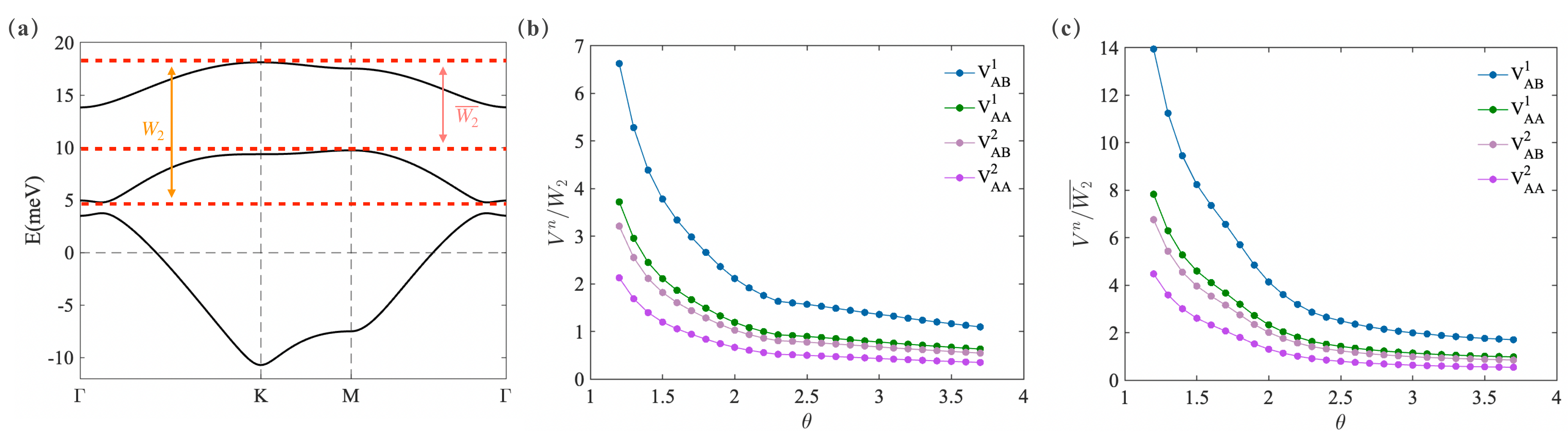}
  \caption{(a) Band structure from continuum model at $\theta = 2.5 ^\circ$.
  (b)/(c) The relation between the density-density interaction and twist angle, the definition of  $W_2$ and $\overline{W_2}$  are shown in the (a). We note the strength of density-density interaction surpass the energy separation between 1st band and 2nd moire bands from 1.2$^\circ $ to  3.7$^\circ$.
  }\label{fig:Sm_w2}
\end{figure}

\subsection{The parameter of the two band model}
We obtain these parameters directly from the Wannier function of our two-band model, and we additionally select a specific gauge to ensure that the nearest-neighbor hopping becomes a real quantity. It's important to highlight that the substantial imaginary component in the next-nearest-neighbor hopping term contributes to the emergence of the topological gap.

\begin{table}[h]
    \caption{
    The parameter of the 2band model,  the screening length is chosen to be $r_0=2a_m$ here and $\epsilon =10$. And all term are presented with the units of meV. $t^{c1}_{BB}$ is the correlated hopping.
    And we additionally select a specific gauge to ensure that the nearest-neighbor hopping becomes a real quantity.}
    \resizebox{1\linewidth}{!}{
    \begin{tabular}{llllllllllllll}
    \multicolumn{1}{c}{$\theta$} & \multicolumn{1}{c}{$t^1_{AB}$} & \multicolumn{1}{c}{$t^1_{AA}$} & \multicolumn{1}{c}{$t^2_{AB}$} & \multicolumn{1}{c}{$t^3_{AB}$}& \multicolumn{1}{c}{$t^2_{AA}$}& \multicolumn{1}{c}{$t^3_{AA}$}& \multicolumn{1}{c}{$U_A$} & \multicolumn{1}{c}{$V^1_{AB}$} & \multicolumn{1}{c}{$V^1_{AA}$} & \multicolumn{1}{c}{$V^2_{AB}$} & \multicolumn{1}{c}{$J^1_{AB}$} & \multicolumn{1}{c}{$J^1_{AA}$} & \multicolumn{1}{c}{$t^{c1}_{BB}$} \\ \hline
    1.2                         & -0.249                & -0.014 + 0.018i & 0.008 - 0.001i     & -0.001 + 0.000i &0.000 + 0.000i &0.000 + 0.000i  & 41.160               & 9.721                    & 5.458                    & 4.7122                    & 0.0349                    & 0.0005        & -0.018 - 0.029i         \\ 
    1.3                         & -0.344                & -0.026 + 0.035i & 0.016 + 0.000i     & -0.003 + 0.000i &0.000 + 0.000i &0.000 + 0.000i  & 41.985               & 10.591           & 5.929                    & 5.116                 & 0.0622                    & 0.0016                    & -0.033 - 0.053i         \\ 
    1.4                         & -0.439                & -0.040 + 0.053i & 0.013 - 0.002i     & -0.004 + 0.003i &0.010 - 0.001i &-0.003 + 0.005i  & 43.709               & 11.456           & 6.390                    & 5.512                 & 0.0729                    & 0.0018                    & -0.045 - 0.051i         \\ 
    1.5                         & -0.574                & -0.068 + 0.094i & 0.049 + 0.000i     & -0.011 + 0.000i &0.000 + 0.000i &-0.001 - 0.002i  & 43.040               & 12.350           & 6.893                    & 5.942                 & 0.154                  & 0.0085                    & -0.077 - 0.125i         \\ 
    1.6                         & -0.703                & -0.098 + 0.139i & 0.076 + 0.000i     & -0.019 + 0.000i &0.000 + 0.000i &-0.001 - 0.005i  & 43.362               & 13.232           & 7.386                    & 6.364                 & 0.219                  & 0.0155                    & -0.101 - 0.170i         \\ 
    1.7                         & -0.840                & -0.133 + 0.194i  & 0.111 + 0.000i    & -0.030 + 0.000i &-0.003 + 0.000i&-0.003 - 0.009i   & 43.617               & 14.110           & 7.885                    & 6.792                 & 0.296                  & 0.0250                    & -0.120 - 0.215i         \\ 
    1.8                         & -0.981                & -0.172 + 0.259i  & 0.153 + 0.000i    & -0.043 + 0.000i &-0.006 + 0.000i&-0.004 - 0.015i   & 43.854               & 14.981           & 8.390                    & 7.227                 & 0.382                  & 0.0366                    & -0.130 - 0.257i         \\ 
    1.9                         & -1.125                & -0.215 + 0.333i  & 0.202 + 0.000i    & -0.059 + 0.000i &-0.011 + 0.000i&-0.007 - 0.024i   & 44.104               & 15.844           & 8.897                    & 7.666                 & 0.472                  & 0.0498                    & -0.129 - 0.292i         \\ 
    2.0                         & -1.267                & -0.260 + 0.415i  & 0.258 + 0.000i    & -0.080 - 0.001i &-0.019 + 0.000i&-0.010 - 0.035i   & 44.387               & 16.697           & 9.406                    & 8.110                 & 0.566                  & 0.0638                    & -0.112 - 0.317i         \\ 
    2.1                         & -1.408                & -0.307 + 0.505i  & 0.320 + 0.000i    & -0.101- 0.001i &-0.028 + 0.000i &-0.013 - 0.049i  & 44.711               & 17.538           & 9.914                    & 8.558                 & 0.658                  & 0.0781                    & -0.077 - 0.327i         \\ 
    2.2                         & -1.544                & -0.355 + 0.601i  & 0.389 + 0.000i    & -0.126- 0.002i &-0.038 + 0.000i &-0.016 - 0.066i  & 45.077               & 18.369           & 10.421                  & 9.0092                & 0.749                 & 0.0923                    & -0.022 - 0.319i         \\ 
    2.3                         & -1.674                & -0.403 + 0.704i  & 0.463 + 0.000i    & -0.152- 0.001i &-0.050 + 0.000i &-0.019 - 0.086i  & 45.448               & 19.185           & 10.923                  & 9.459               & 0.838                    & 0.106                   & 0.056 - 0.288i          \\ 
    2.4                         & -1.798                & -0.451 + 0.813i  & 0.545 + 0.000i    & -0.178- 0.001i &-0.063 + 0.000i &-0.022 - 0.108i  & 45.929               & 20.005           & 11.427                  & 9.920                   & 0.924                    & 0.120                    & 0.156 - 0.234i          \\ 
    2.5                         & -1.944                & -0.507 + 0.923i  & 0.609 + 0.000i    & -0.206- 0.004i &-0.085 + 0.000i &-0.028 - 0.119i  & 47.116               & 20.941           & 11.941                  & 10.332               & 0.964                    & 0.142                    & 0.134 - 0.374i          \\ 
    2.6                         & -2.067                & -0.563 + 1.041i & 0.688 + 0.001i     & -0.233- 0.004i &-0.104+ 0.003i  &-0.032 - 0.140i & 47.852               & 21.792           & 12.446                  & 10.776               & 1.023                    & 0.159                    & 0.220 - 0.364i          \\ 
    2.7                         & -2.186                & -0.620 + 1.163i & 0.769 + 0.000i     & -0.260- 0.005i &-0.126+ 0.004i  &-0.035 - 0.161i & 48.642               & 22.644           & 12.950                  & 11.217               & 1.073                    & 0.177                    & 0.317 - 0.350i          \\ 
    2.8                         & -2.301                & -0.678 + 1.291i & 0.851 + 0.000i     & -0.288- 0.007i &-0.149+ 0.005i  &-0.037 - 0.183i & 49.480               & 23.500           & 13.453                  & 11.658               & 1.115                    & 0.196                    & 0.425 - 0.332i          \\ 
    2.9                         & -2.409                & -0.738 + 1.423i & 0.934 + 0.000i     & -0.313- 0.010i &-0.175+ 0.007i  &-0.038 - 0.205i & 50.355               & 24.360           & 13.956                  & 12.098               & 1.149                    & 0.217                    & 0.541 - 0.310i          \\ 
    3.0                         & -2.514                & -0.802 + 1.559i & 1.015 + 0.000i     & -0.333- 0.012i &-0.203+ 0.011i  &-0.038 - 0.227i & 51.254               & 25.222           & 14.460                  & 12.536               & 1.175                    & 0.238                    & 0.666 - 0.288i          \\ 
    3.1                         & -2.627                & -0.881 + 1.692i & 1.094 + 0.002i     & -0.346- 0.016i &-0.236+ 0.016i  &-0.031 - 0.228i & 52.858               & 26.242           & 14.959                  & 12.925               & 1.151                    & 0.255                    & 0.735 - 0.393i          \\ 
    3.2                         & -2.710                & -0.934 + 1.843i & 1.177 + 0.000i     & -0.368- 0.018i &-0.265+ 0.019i  &-0.034 - 0.269i & 53.114               & 26.949           & 15.466                  & 13.408               & 1.206                    & 0.287                    & 0.938 - 0.237i          \\
    3.3                         & -2.801                & -1.003 + 1.991i & 1.256 - 0.001i     & -0.386- 0.023i &-0.300+ 0.022i  &-0.030 - 0.291i & 54.069               & 27.813           & 15.969                  & 13.842               & 1.210                    & 0.314                    & 1.086 - 0.210i          \\ 
    3.4                         & -2.887                & -1.074 + 2.143i & 1.334 - 0.001i     & -0.403- 0.029i &-0.337+ 0.025i  &-0.025 - 0.312i & 55.033               & 28.676           & 16.473                  & 14.275               & 1.209                    & 0.343                    & 1.241 - 0.184i          \\ 
    3.5                         & -2.968                & -1.149 + 2.300i & 1.408 + 0.000i     & -0.419- 0.037i &-0.377+ 0.027i  &-0.019 - 0.333i & 56.001               & 29.538           & 16.976                  & 14.708               & 1.202                    & 0.374                    & 1.400 - 0.159i          \\ 
    3.6                         & -3.043                & -1.227 + 2.460i & 1.482 - 0.001i     & -0.433- 0.044i &-0.420+ 0.029i  &-0.012 - 0.354i & 56.970               & 30.397           & 17.479                  & 15.139               & 1.190                    & 0.407                    & 1.565 - 0.135i          \\ 
    3.7                         & -3.114                & -1.309 + 2.625i & 1.554 - 0.001i     & -0.446- 0.053i &-0.466+ 0.032i  &-0.003 - 0.376i & 57.940               & 31.254           & 17.983                  & 15.570               & 1.175                    & 0.442                    & 1.734 - 0.112i          \\
    \end{tabular}}
    \end{table}
\begin{figure}
\includegraphics[width=1.0\columnwidth]{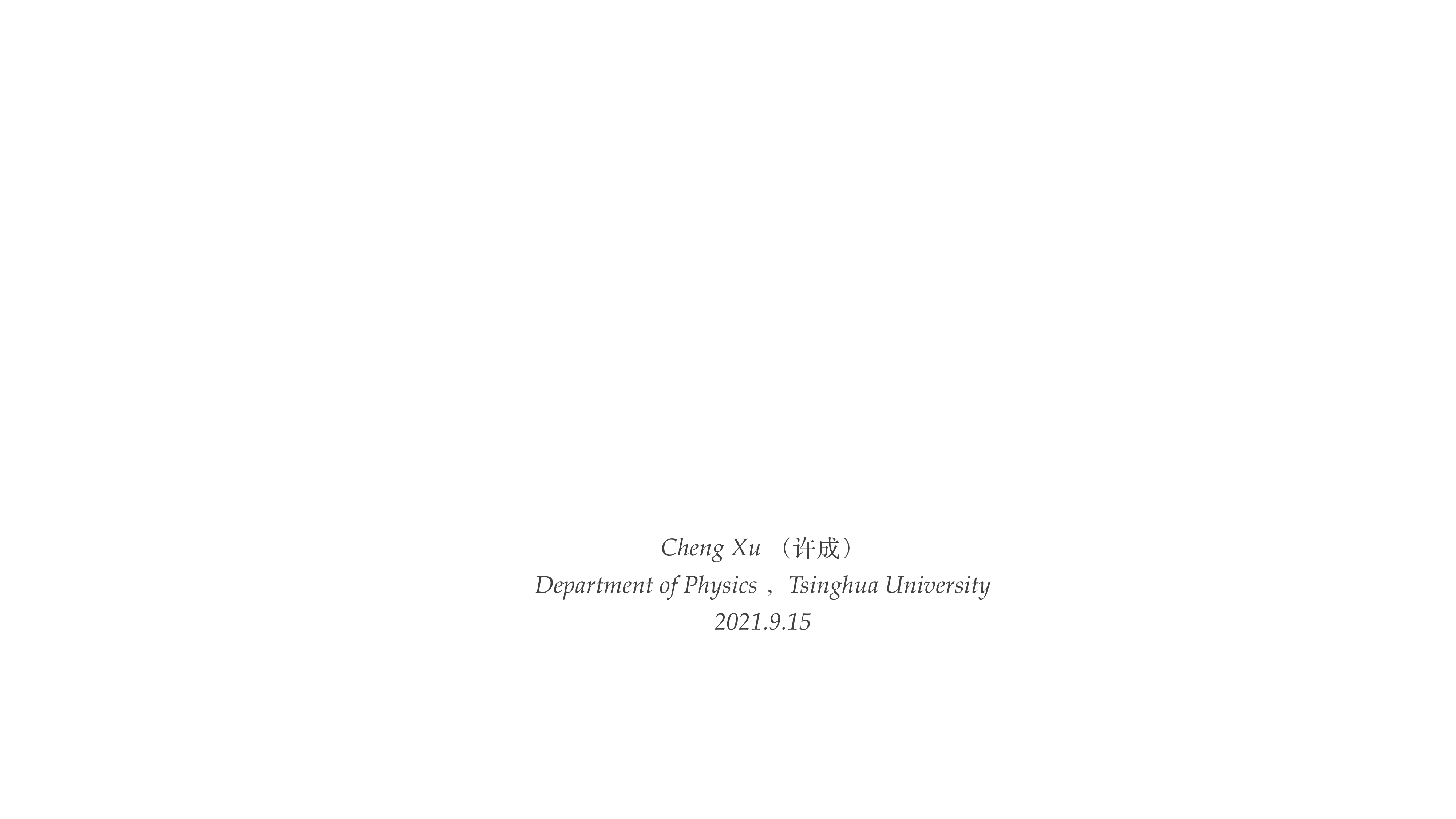}
\caption{ The total momentum counting for the 3-fold degenerate ground states with the geometry: 5×3 }
\label{fig:totalm}
\end{figure}

\begin{figure}
\includegraphics[width=1.0\columnwidth]{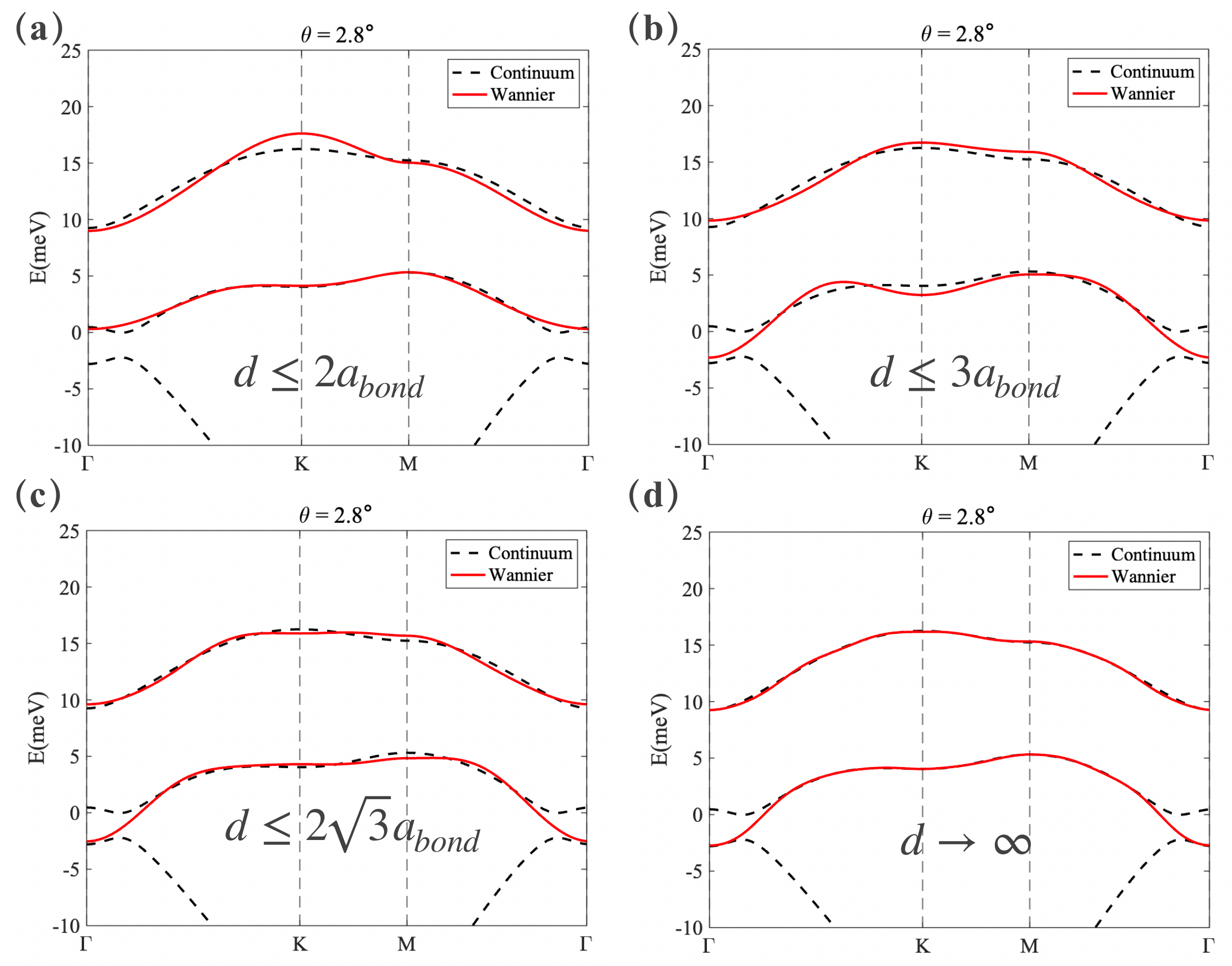}
\caption{ The bands from the two band Wannier function with the different cutoff of hopping at $\theta = 2.8 ^\circ$. d is the distance of the hopping and a$_{bond} $ is the mininal distance between two Wannier Orbits.(a$_{bond}=\frac{a_m}{\sqrt{3}}$) }
\label{fig:twocutoff}
\end{figure}

\end{document}